\documentclass[12pt,draftclsnofoot,journal,onecolumn]{IEEEtran}
\pdfoutput=1
\usepackage{cite}
\usepackage[pdftex]{graphicx}
\usepackage[cmex10]{amsmath}
\usepackage{amssymb}
\usepackage{amsthm}
\usepackage{mathtools}
\usepackage{algorithmic}
\usepackage{array}
\usepackage{mdwmath}
\usepackage{mdwtab}
\usepackage{subcaption}
\usepackage{fixltx2e}
\usepackage{url}
\usepackage{enumerate}

\newtheorem{theorem}{Theorem}
\newtheorem*{theorem*}{Theorem}

\newtheorem{lemma}{Lemma}

\newcounter{customtheoremcounter}
\newtheorem{customtheorem}{Theorem}[customtheoremcounter]

\newcommand{\expected}[1]{\mathbb{E}\left[#1\right]}

\DeclareMathOperator{\Var}{Var}

\newcommand\myatop[2]{\genfrac{}{}{0pt}{}{#1}{#2}} 

\newcommand{\plusser}[2] {
        \put(#1,#2){\begin{picture}(5,5)
                    \put(0,0) {\circle{5}}
                    \put(0,-2) {\line(0,1) {4}}
                    \put(-2,0){\line(1,0) {4}}
                    \end{picture}}
                        }

\begin{document}
\title{Covert Communication Gains from \\Adversary's Ignorance of Transmission Time}
\author{Boulat~A.~Bash,~\IEEEmembership{Member,~IEEE,}
        Dennis~Goeckel,~\IEEEmembership{Fellow,~IEEE,}
        and~Don~Towsley,~\IEEEmembership{Fellow,~IEEE}%
\thanks{B.~A.~Bash is with Raytheon BBN Technologies, Cambridge, MA (email: bbash@bbn.com).}
\thanks{D.~Goeckel is with the Electrical and Computer Engineering Department, University of Massachusetts, Amherst, MA (email: goeckel@ecs.umass.edu).}%
\thanks{D.~Towsley is with the College of Information and Computer Sciences, University of Massachusetts, Amherst, MA (email: towsley@cs.umass.edu).}%
\thanks{This research was sponsored by the National Science Foundation under 
  grants CNS-1018464, CNS-0964094, and ECCS-1309573. 
  B.~A.~Bash acknowledges financial support from Raytheon BBN Technologies and
  DARPA under contract number HR0011-16-C-0111.}
\thanks{An abridged version of this manuscript was presented at the International Symposium on Information Theory (ISIT) 2014 \cite{bash14timingisit}.}
\thanks{This work has been submitted to the IEEE for possible publication.  Copyright may be transferred without notice, after which this version may no longer be accessible.}}

\IEEEspecialpapernotice{UMass Technical Report UM-CS-2014-001}

\maketitle
\vspace{-0.40in}
\begin{abstract}
The recent \emph{square root law} (SRL) for covert communication demonstrates
  that Alice can reliably transmit $\mathcal{O}(\sqrt{n})$ bits to
  Bob in $n$ uses of an additive white Gaussian noise (AWGN) channel while
  keeping ineffective any detector employed by the adversary; conversely,
  exceeding this limit either results in detection by the adversary with high 
  probability or non-zero decoding error probability at Bob.
This SRL is under the assumption that the adversary knows \emph{when} Alice
  transmits (if she transmits); however, in many operational scenarios
  he does not know this.
Hence, here we study the impact of the adversary's ignorance of 
  the time of the communication attempt.
We employ a slotted AWGN channel model with $T(n)$ slots each containing 
  $n$ symbol periods, where Alice may use a single slot out of $T(n)$.
Provided that Alice's slot selection is secret, the adversary needs to monitor 
  all $T(n)$ slots for possible transmission.
We show that this allows Alice to reliably transmit 
  $\mathcal{O}(\min\{\sqrt{n\log T(n)},n\})$ bits to Bob (but no more) while 
  keeping the adversary's detector ineffective.
To achieve this gain over SRL,
Bob does not have to know the time of transmission provided
  $T(n)<2^{c_{\rm T}n}$, $c_{\rm T}=\mathcal{O}(1)$.
\end{abstract}

\IEEEpeerreviewmaketitle

\section{Introduction}
\label{sec:introduction}
Recent revelations of massive surveillance 
  programs \cite{bbc14snowden} have emphasized the
  need for secure communication systems that do not just protect
  the content of the user's message from being decoded, but prevent
  the detection of its transmission in the first place.
Indeed, encrypted data or even just the transmission of a signal can arouse 
  suspicion, and even the most theoretically robust cryptographic security 
  scheme can be defeated by a determined adversary using non-computational 
  methods such as side-channel analysis.
Covert, or \emph{low probability of detection/intercept} (LPD/LPI) communication
  capability is thus very important in extremely sensitive situations 
  (e.g., during military operations or for organization of civil unrest).

In the covert communication scenario, Alice transmits a message to Bob over a 
  noisy channel while the adversary, warden Willie, attempts to detect her 
  transmission.
The channel from Alice to Willie is also subject to noise.
Thus, while Alice transmits low-power covert signals to Bob, Willie attempts to 
  classify these signals as either noise on his channel or signals from Alice.
We recently showed that the \emph{square root law} (SRL) governs 
  covert communication: provided that Alice and Bob pre-share a secret of
  sufficient length,
  she can reliably transmit $\mathcal{O}(\sqrt{n})$ bits
  to Bob in $n$ uses of an additive white Gaussian noise (AWGN) channel while
  keeping ineffective any detector employed by Willie.
Conversely, attempting to transmit more that $\mathcal{O}(\sqrt{n})$ bits either
  results in detection by Willie with probability one or non-zero decoding 
  error probability at Bob as $n\rightarrow\infty$
  \cite{bash12sqrtlawisit,bash13squarerootjsac}.
Follow-on work has addressed the size of the pre-shared
  secret \cite{che13sqrtlawbscisit, bloch15covert},
  the optimal constant hidden by the asymptotic (big-$\mathcal{O}$) notation
  \cite{bloch15covert, wang15covert}, network aspects of
  covert communication 
  \cite{kadhe14sqrtlawmultipathisit,soltani14netlpdallerton},
  extensions of the SRL to quantum channels where the adversary is only limited
  by the laws of quantum mechanics \cite{bash13quantumlpdisit,
  bash15covertbosoniccomm,azadeh15quantumcovertarxiv},
  and, finally, practical limitations on the adversary that improve 
  the covert throughput to beyond the SRL limit 
  \cite{lee15posratecovertjstsp,hou14isit,han14reliabilitysecrecy,
  che14channeluncertainty,sobers15jammer-asilomar}.

Studies of covert communication up to now assume that Willie knows when
  Alice may start to transmit (if she does).
However, in many practical scenarios (e.g., delay-constrained communication), 
  Alice's message may have to be short relative to the total time available 
  to transmit it (e.g.,~a few seconds out of the day when both Alice and Bob are
  available).
Additionally, Willie might not know when the communication starts (e.g., 
  a possible transmission time can be secretly pre-arranged in advance).
This forces Willie to monitor a much longer time period than the duration
  of Alice's transmission, limiting his capabilities.
Here we show how Alice can 
  leverage Willie's ignorance of her transmission time
  to transmit significant additional information covertly to Bob. 

\vspace{0.15in}
\begin{figure}[h]
\begin{center}
\begin{picture}(85,13)
\put(0,5.5){\framebox(14.8,4){\parbox{15\unitlength}{\small \centering slot $1$}}}
\put(0,10){\makebox(14.8,5){$\overbrace{\hspace{18mm}}^{n}$}}
\put(15,5.5){\framebox(14.85,4){\parbox{15\unitlength}{\small \centering slot $2$}}}
\put(15,10){\makebox(14.8,5){$\overbrace{\hspace{18mm}}^{n}$}}
\put(30,5.5){\framebox(12.35,4){\parbox{12.4\unitlength}{\small \centering $\cdots$}}}
\put(42.5,5.5){\framebox(14.85,4){\parbox{15\unitlength}{\small \centering slot $t_{\rm A}$}}}
\put(42.5,10){\makebox(14.8,5){$\overbrace{\hspace{18mm}}^{n}$}}
\put(57.5,5.5){\framebox(12.3,4){\parbox{12.4\unitlength}{\small \centering $\cdots$}}}
\put(70,5.5){\framebox(14.8,4){\parbox{15\unitlength}{\small \centering slot $T(n)$}}}
\put(70,10){\makebox(15,5){$\overbrace{\hspace{18mm}}^{n}$}}
\put(0,0){\makebox(85,5){$\underbrace{\hspace{102mm}}_{nT(n)\text{~total symbol periods}}$}}
\put(57.5,14){\vector(-1,-2){3}}
\put(58,14){\makebox(20,3){\scriptsize Slot used by Alice and Bob}}
\end{picture}
\end{center}
\vspace{-0.20in}
\caption{Slotted channel:  each of the $T(n)$ slots contains $n$ symbol periods.
  Alice and Bob use slot $t_{\rm A}$ to communicate.}
\label{fig:timing}
\vspace{-0.10in}
\end{figure}
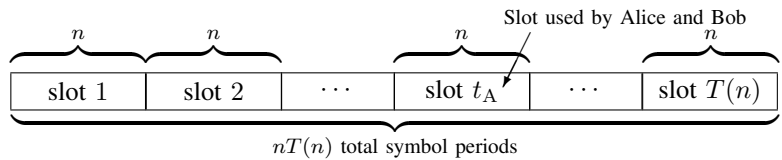

In our scenario, Alice communicates to Bob over an AWGN channel.
Willie also has an AWGN channel from Alice.
Unlike the setting in 
  \cite{bash12sqrtlawisit,bash13squarerootjsac}, the 
  channel is slotted, as shown in Figure \ref{fig:timing}.
Each of $T(n)$ slots contains $n$ symbol periods, where $T(n)$ is
  an increasing function of $n$.
If Alice used all $nT(n)$ symbol periods for transmission, then, by 
  the SRL in \cite{bash12sqrtlawisit,bash13squarerootjsac}, she 
  could reliably transmit $\mathcal{O}(\sqrt{nT(n)})$ covert 
  bits to Bob.
However, to model a practical scenario where Alice is
  constrained to a short message relative to the total available transmission 
  time, her communication is restricted to a single slot $t_{\rm A}$ which
  is kept secret from Willie.
While this certainly reduces the amount of transmissible data,
  a natural question is whether an improvement can be made over 
  a na\"{i}ve application of the SRL
  \cite{bash12sqrtlawisit,bash13squarerootjsac}, which
  allows Alice to reliably transmit $\mathcal{O}(\sqrt{n})$ covert bits 
  in this scenario.
Here we demonstrate that Alice can transmit
  $\mathcal{O}\left(\min\{\sqrt{n \log T(n)},n\}\right)$ bits reliably on this 
  channel while maintaining arbitrarily low probability of detection by Willie.
Conversely, we show that the transmission of $\omega(\sqrt{n\log T(n)})$ 
  bits\footnote{Throughout this paper we employ asymptotic notation 
  \cite[Ch.~3.1]{clrs2e} where $f(n)=\mathcal{O}(g(n))$ denotes an asymptotic
  upper bound on $f(n)$ (i.e., there exist constants $m,n_0>0$ such 
  that $0\leq f(n)\leq m g(n)$ for all $n\geq n_0$), 
  $f(n)=o(g(n))$ denotes an upper bound on $f(n)$ that is not 
  asymptotically tight (i.e., for any constant $m>0$, there exists constant 
  $n_0>0$ such that $0\leq f(n)<m g(n)$ for all $n\geq n_0$),
  and $f(n)=\omega(g(n))$ denotes a lower bound on $f(n)$ that is not 
  asymptotically tight (i.e., for any constant $m>0$, there exists constant 
  $n_0>0$ such that $0\leq m g(n)<f(n)$ for all $n\geq n_0$).}
  either results in Alice being detected with high probability or unreliable 
  communication.

The improvement stems from Willie not knowing $t_{\rm A}$ and being forced 
  to monitor all $T(n)$ slots.
When Willie knows which slot Alice might use, by Theorem 2 in
  \cite{bash12sqrtlawisit,bash13squarerootjsac}
  she can be detected if she transmits 
  more than $\mathcal{O}(\sqrt{n})$ bits, since it is improbable 
  that Willie's observations will look like AWGN noise that he expects.
His optimal detector is a threshold on the power observed in slot $t_{\rm A}$
  \cite{sobers15jammer-asilomar}.
However, if Willie does not know $t_{\rm A}$, he has to examine
  all $T(n)$ slots.
Effectively, Willie's test statistic is the maximum slot power
  (see the remark following the proof of Theorem \ref{th:achievability}).
When only noise is observed, the average maximum slot power is substantially 
  higher than average power in any single slot.
Hence, to avoid false alarms, Willie's threshold when he does not know
  $t_{\rm A}$ must also be greater than when he knows it.
This allows Alice to transmit additional covert information.

Our main result is stated formally as follows:

\begin{theorem*}
Suppose the channel between Alice and each of Bob and Willie experiences 
  independent additive white Gaussian noise (AWGN) with constant power 
  $\sigma_{\rm b}^2>0$ and $\sigma_{\rm w}^2>0$, respectively,
  and that Alice's transmitter is subject to the average power constraint
  $P_{\max}\in(0,\infty)$.
Also suppose that, if Alice chooses to transmit, she uses one of the $T(n)$ 
  slots chosen randomly.
Each slot contains $n$ symbol periods, where $T(n)=\omega(1)$.
Then, for any $\epsilon>0$, Alice can reliably transmit 
  $\mathcal{O}\left(\min\{\sqrt{n\log T(n)},n\}\right)$ bits to Bob
  in a selected slot while maintaining a probability of detection error
  by Willie greater than $\frac{1}{2}-\epsilon$. 
Conversely, if Alice tries to transmit $\omega(\sqrt{n \log T(n)})$ bits using
  $n$ consecutive symbol periods, either Willie detects her with arbitrarily low
  probability of error or Bob cannot decode her message with arbitrary low 
  probability of decoding error as $n\rightarrow\infty$.
\end{theorem*}

As in \cite{bash12sqrtlawisit, bash13squarerootjsac}, covert communication
  requires that Alice and Bob possess a common randomness resource.
This corresponds to a secret codebook\footnote{The requirement of a pre-shared 
  secret was shown to be unnecessary \cite{che13sqrtlawbscisit, bloch15covert} 
  for the standard SRL in \cite{bash12sqrtlawisit, bash13squarerootjsac} if Bob
  has a better channel than Willie; \cite{bloch15covert} was extended 
  \cite{arumugam16async} to the asynchronous scenario described in this paper 
  while it was in review.} needed in our proofs that is shared between
  Alice and Bob prior to communication, analogous to the one-time pad
  in the information-theoretic analysis of encryption \cite{shannon49sec}.
This follows ``best practices'' in security system design as the security of
  the covert communication system depends only on the shared secret 
  \cite{menezes96HAC}.
Remarkably, we demonstrate that the \emph{multiplicative}
  increase (by a factor of $\sqrt{\log T(n)}$) in the number of covert bits 
  that Alice can transmit reliably to Bob does not require Bob to know
  the timing of the transmission if $T(n)<2^{c_{\rm T}n}$, where $c_{\rm T}>0$ 
  is a constant;
  to realize the $\sqrt{\log T(n)}$ gain when $T(n)\geq 2^{c_{\rm T}n}$ only an 
  additive expense of  an extra $\log T(n)$ secret bits is needed to indicate
  to Bob the slot employed by Alice.
Thus, at most $\log T(n)$ secret bits are required in excess of those needed 
  to enable the SRL on a single $n$-symbol slot.
Timing is therefore a very useful resource for covert communication.
It also necessitates a vastly different analysis than that
  in \cite{bash12sqrtlawisit,bash13squarerootjsac}.
Specifically, the relative entropy based bounds on the probability
  of detection error used in \cite{bash12sqrtlawisit,bash13squarerootjsac}
  are too loose to yield our achievability results, and
  we thus have to apply other techniques from mathematical statistics.

After introducing our slotted channel model in Section 
  \ref{sec:prerequisites}, we prove the achievability and the converse
  in Sections \ref{sec:achievability} and \ref{sec:converse},
  respectively.
We discuss the relationship of our paper to other work in covert communication
  in Section \ref{sec:discussion} and conclude in Section \ref{sec:conclusion}.

\section{Prerequisites}
\label{sec:prerequisites}

\subsection{Channel Model}
\begin{figure}[h]
\vspace{0.0in}
\begin{center}
\begin{picture}(85,42)
\put(33,38){\framebox(14,4){\parbox{14\unitlength}{\small \centering secret}}}
\put(7,40){\vector(0,-1){3}}
\put(73,40){\vector(0,-1){3}}
\put(7,40){\line(1,0){26}}
\put(47,40){\line(1,0){26}}
\put(0,31){\framebox(15,6){Alice}}
\put(15,34){\vector(1,0){38.25}}
\put(16,26.5){\makebox(25,5){\small $...,0,\underbrace{f_1,...,f_n},0,...$}}
\put(17,21.5){\makebox(25,5){\small $\text{transmit in slot }t_{\rm A}$}}
\put(47,34){\circle*{1}}
\put(47,34){\vector(0,-1){16.5}}
\plusser{47}{15}
\put(49.5,15){\vector(1,0){12.5}}
\put(62,12){\framebox(15,6){Willie}}
\put(45,2){\makebox(58,12){\parbox{53\unitlength}{\small \centering decide: $\mathbf{Z}_{\rm w}$ or not?}}}
\put(32,15){\vector(1,0){12.5}}
\put(25,12){\makebox(7,6){$Z_i^{({\rm w})}$}}
\put(52.75,21.5){\makebox(7,6){$Z_i^{({\rm b})}$}}
\plusser{55.75}{34}
\put(55.75,27.5){\vector(0,1){4}}
\put(58.25,34){\vector(1,0){6.75}}
\put(65,31){\framebox(15,6){Bob}}
\put(63,25){\makebox(25,5){\small decode $f_1,\ldots,f_n$}}
\end{picture}
\end{center}
\vspace{-0.35in}
\caption[System framework for covert communication over slotted AWGN channels]{System framework: Alice and Bob share a secret before transmission.
  If Alice chooses to transmit, she encodes information into a vector of real 
  symbols $\mathbf{f}=\{f_i\}_{i=1}^n$ and uses random slot $t_{\rm A}$ to send 
  it on an AWGN channel to Bob (to ensure reliable decoding $t_{\rm A}$ is 
  secretly shared with Bob before the transmission if $T(n)\geq2^{c_{\rm T}n}$, 
  where $c_{\rm T}$ is a constant). 
  Upon observing the channel from Alice, Willie has to classify his vector 
  of readings $\mathbf{Y}_{\rm w}$ as either an AWGN vector
  $\mathbf{Z}_{\rm w}=\{Z^{({\rm w})}_i\}_{i=1}^{nT(n)}$ or a vector that 
  contains a slot with transmissions corrupted by AWGN.}
\label{fig:framework}
\vspace{-0.10in}
\end{figure}
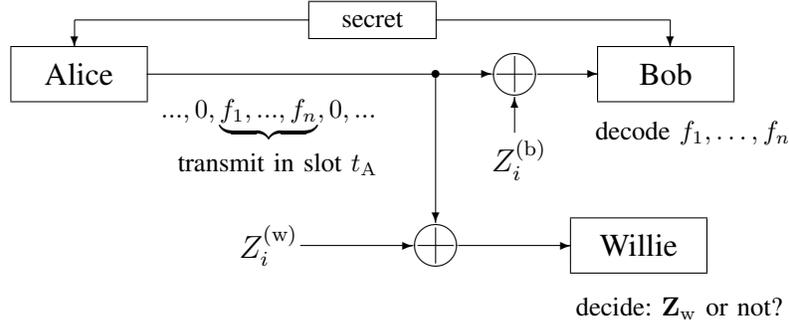

We use the discrete-time slotted AWGN channel model with real-valued symbols 
  depicted in Figures \ref{fig:timing} and \ref{fig:framework}.
The channel has $T(n)$ slots, each containing $n$ symbol periods.
Alice selects slot $t_{\rm A}$ uniformly at random prior to transmission.
If Alice chooses to transmit, she sends a vector of $n$ 
  real-valued symbols $\mathbf{f}=\{f_i\}_{i=1}^n$ during slot $t_{\rm A}$.
The AWGN on Bob's channel is described by an independent and identically 
  distributed (i.i.d.) sequence $\{Z^{({\rm b})}_i\}_{i=1}^{nT(n)}$ of $nT(n)$ 
  zero-mean Gaussian random variables with variance $\sigma_{\rm b}^2$ (i.e., 
  $Z^{({\rm b})}_i\sim \mathcal{N}(0,\sigma_{\rm b}^2)$).
Bob receives $\mathbf{Y}_{\rm b}=\{\mathbf{Y}_{\rm b}(t)\}_{t=1}^{T(n)}$ where
  $\mathbf{Y}_{\rm b}(t)=[Y^{({\rm b})}_{(t-1)n+1},\ldots,Y^{({\rm b})}_{tn}]$ is a vector of
  observations collected during slot $t$.
If Alice transmits during slot $t_{\rm A}$,
  $Y^{({\rm b})}_{(t_{\rm A}-1)n+i}=f_{i}+Z^{({\rm b})}_{(t_{\rm A}-1)n+i}$.
For any slot that is not used for transmission,
  $Y^{({\rm b})}_{(t-1)n+i}=Z^{({\rm b})}_{(t-1)n+i}$ (this includes all slots 
  $\{t:t\neq t_{\rm A}\}$, and slot $t_{\rm A}$ when Alice does not transmit).

Similarly, Willie observes 
  $\mathbf{Y}_{\rm w}=\{\mathbf{Y}_{\rm w}(t)\}_{t=1}^{T(n)}$ where 
  $\mathbf{Y}_{\rm w}(t)=[Y^{({\rm w})}_{(t-1)n+1},\ldots,Y^{({\rm w})}_{tn}]$ is a vector of
  observations collected during slot $t$.
The AWGN on Willie's channel is described by an i.i.d.~sequence 
  $\{Z^{({\rm w})}_i\}_{i=1}^{nT(n)}$ of $nT(n)$ zero-mean Gaussian random variables 
  with variance $\sigma_{\rm w}^2$ (i.e., $Z^{({\rm w})}_i\sim\mathcal{N}(0,\sigma_{\rm w}^2)$).
If Alice transmits during slot $t_{\rm A}$, 
  $Y^{({\rm w})}_{(t_{\rm A}-1)n+i}=f_{i}+Z^{({\rm w})}_{(t_{\rm A}-1)n+i}$.
For any slot that is not used for transmission,
  $Y^{({\rm w})}_{(t-1)n+i}=Z^{({\rm w})}_{(t-1)n+i}$ (again, this includes all slots 
  $\{t:t\neq t_{\rm A}\}$, and slot $t_{\rm A}$ when Alice does not transmit).
Willie does not know $t_{\rm A}$ and has to perform a statistical hypothesis test on
  his entire set of observations $\mathbf{Y}_{\rm w}$ to determine whether Alice is
  transmitting.

\subsection{Hypothesis Testing}
\label{sec:hyptest}
Willie performs a statistical hypothesis test \cite{lehmann05stathyp} on 
  $\mathbf{Y}_{\rm w}$, where the null hypothesis $H_0$ is that Alice does not 
  transmit and each sample is an i.i.d.~realization of AWGN 
  $Y^{({\rm w})}_i\sim \mathcal{N}(0,\sigma_{\rm w}^2)$.
The alternate hypothesis $H_1$ is that Alice transmits,
  and the samples from one of the slots come from a different distribution.
The rejection of $H_0$ when it is true is a false alarm (FA)
  and the acceptance of $H_0$ when it is false is a missed detection (MD).
The lower bound on the sum of the probabilities of these events 
  $\mathbb{P}_{\rm FA}+\mathbb{P}_{\rm MD}$ characterizes the necessary
  trade-off between the false alarms and missed detections in a hypothesis test.
As in \cite{bash12sqrtlawisit,bash13squarerootjsac}, Alice desires
  $\mathbb{P}_{\rm FA}+\mathbb{P}_{\rm MD}\geq 1-\epsilon$ for an arbitrary 
  choice of $\epsilon>0$, ensuring that any hypothesis test Willie
  constructs is ineffective.\footnote{Willie's probability of error is
  $\mathbb{P}_{\rm e}^{({\rm w})}=\pi_0\mathbb{P}_{\rm FA}+\pi_1\mathbb{P}_{\rm MD}$, 
  where $\pi_0$ and $\pi_1$ are the prior probabilities of hypotheses
  $H_0$ and $H_1$.
  A random guess of Alice's transmission state and ``uninformative'' priors 
  $\pi_0=\pi_1=\frac{1}{2}$ yield $\mathbb{P}_{\rm e}^{({\rm w})}=\frac{1}{2}$.
  By lower-bounding $\mathbb{P}_{\rm FA}+\mathbb{P}_{\rm MD}\geq 1-\epsilon$,
  Alice ensures that, for uninformative priors, any detector Willie employs 
  can only be slightly better than a random guess.
  Our scaling results apply when $\pi_0\neq\pi_1$ as well, since
  $\mathbb{P}_{\rm e}^{({\rm w})}\geq\min(\pi_0,\pi_1)\left(\mathbb{P}_{\rm FA}+\mathbb{P}_{\rm MD}\right)$.}

\section{Achievability}
\label{sec:achievability}

\subsection{Proof Idea and Preliminaries}
We proved the achievability theorems in 
  \cite{bash12sqrtlawisit,bash13squarerootjsac} by upper-bounding the relative
  entropy \cite[Ch.~10]{cover02IT} between the distributions of 
  Willie's sequence of observations $\mathbf{Y}_{\rm w}$ corresponding to 
  hypotheses $H_0$ and $H_1$.
Here we take a different approach by explicitly analyzing
  Willie's optimal detector assuming that his only unknowns are:
\begin{enumerate}[a)]
\item slot $t_{\rm A}$ chosen uniformly at random from $\{1,\ldots,T(n)\}$ by 
  Alice for transmission (if she transmits), as 
  depicted in Figures \ref{fig:timing} and \ref{fig:framework}; and
\item a secret shared between Alice and Bob prior to the potential 
  transmission.
\end{enumerate}
Thus, Willie is given Alice's channel input distribution, 
  the distribution of the AWGN on his channel from Alice, 
  and the slot boundaries depicted in Figure \ref{fig:timing}.
This effectively provides Willie with a complete statistical model of 
  observations $\mathbf{Y}_{\rm w}$, allowing him to construct the likelihood
  functions $f_0(\mathbf{Y}_{\rm w})$ and $f_1(\mathbf{Y}_{\rm w})$ under 
  hypotheses $H_0$ and $H_1$, respectively.
Alice's transmission state is binary (either she transmits or she does not) and
  the optimal detector for Willie is the likelihood ratio test (LRT) by
  the Neyman--Pearson lemma~\cite[Ch.~3.2 and 13.1]{lehmann05stathyp}.
Here we determine what it takes for Alice to ensure that Willie's optimal
  detector performs only slightly better than a random guess of her transmission
  state, and how much data she can reliably transmit to Bob in this manner.

The LRT compares the likelihood ratio
  $\Lambda(\mathbf{Y}_{\rm w})=\frac{f_1(\mathbf{Y}_{\rm w})}{f_0(\mathbf{Y}_{\rm w})}$ to 
  a threshold $\tau(n)$.
$H_0$ or $H_1$ is chosen based on whether $\Lambda(\mathbf{Y}_{\rm w})$ is smaller or 
  larger than $\tau(n)$ (if it equals the threshold, a random decision is made):
\begin{align}
\label{eq:hyptest_s}\Lambda(\mathbf{Y}_{\rm w})&\mathop{\lessgtr}^{H_0}_{H_1}\tau(n).
\end{align}
The LRT statistic $\Lambda(\mathbf{Y}_{\rm w})$ is a function of the sequence of
  observations $\mathbf{Y}_{\rm w}$, and, as such, is a random variable.
Per its definition in Section \ref{sec:prerequisites}, $\mathbf{Y}_{\rm w}$ 
  is parameterized by the slot length $n$ and which hypothesis is true (that is,
  Alice's transmission state).
Let $\Lambda_s^{(n)}\equiv\Lambda(\mathbf{Y}_{\rm w})$
  where $s\in\{0,1\}$ indicates the true hypothesis ($H_0$ or $H_1$). 
Since one-to-one transformations of both sides in \eqref{eq:hyptest_s} do not 
  affect the performance of the test, we analyze a detector that is equivalent
  to the one defined in \eqref{eq:hyptest_s} but employs 
  the test statistic $L^{(n)}_s\equiv g_n\left(\Lambda_s^{(n)}\right)$,
where $g_n(x)$ is a one-to-one function defined later.
Denote by $K^{(n)}\xrightarrow{\mathcal{P}}Q$ and
  $K^{(n)}\xrightarrow{\mathcal{D}}Q$ convergence of random variable
  $K^{(n)}$ to random variable $Q$ in probability and in distribution,
  respectively.
The following lemma establishes sufficient conditions for the covertness of
  Alice's transmission:
\begin{lemma}
\label{lemma:LRT_bound}
If the LRT statistic is described by random variables:
\begin{align}
L^{(n)}_0&=S^{(n)}+V^{(n)}_0\text{~when~}H_0\text{~is true}\\
L^{(n)}_1&=S^{(n)}+V^{(n)}_1\text{~when~}H_1\text{~is true},
\end{align}
where $V^{(n)}_0\xrightarrow{\mathcal{P}}0$ and 
  $V^{(n)}_1\xrightarrow{\mathcal{P}}0$, as well as 
  $S^{(n)}\xrightarrow{\mathcal{D}}Z$ with $Z\sim\mathcal{N}(0,1)$, then 
  $\mathbb{P}_{\rm FA}+\mathbb{P}_{\rm MD}\geq1-\epsilon$ for any 
  $\epsilon>0$ and a sufficiently large $n$.
\end{lemma}
\begin{IEEEproof}
  See Appendix \ref{app:convergence}.
\end{IEEEproof}

In order to employ Lemma \ref{lemma:LRT_bound}, we require a one-to-one
  function $g_n(x)$ that re-scales $\Lambda^{(n)}_s$ so that 
  the convergence conditions hold.
In the proofs that follow, we show that
  $\Lambda^{(n)}_s=\frac{1}{T(n)}\sum_{t=1}^{T(n)}U_t^{(n)}$,
  where $\{U_t^{(n)}\}_{t=1}^{T(n)}$ is a sequence of $T(n)$ independent 
  random variables, each corresponding to a slot of length $n$.
Since Alice is limited to slot $t_{\rm A}$ for a potential transmission, 
  the only random variable in $\{U_t^{(n)}\}_{t=1}^{T(n)}$ that is
  distributed differently under each hypothesis is $U_{t_{\rm A}}^{(n)}$.
We thus denote by $U_{t_{\rm A}}^{(n,0)}$ and $U_{t_{\rm A}}^{(n,1)}$ the random
  variable corresponding to slot $t_{\rm A}$ under $H_0$ and $H_1$, 
  respectively.
Regardless of Alice's transmission state, 
  $\{U_t^{(n)}\}_{\myatop{t=1}{t\neq t_{\rm A}}}^{T(n)}$ is
  an i.i.d.~sequence, with $U_{t_{\rm A}}^{(n,0)}$ distributed identically
  to the elements of $\{U_t^{(n)}\}_{\myatop{t=1}{t\neq t_{\rm A}}}^{T(n)}$.
Therefore,
\begin{align}
\Lambda^{(n)}_s&=\frac{1}{T(n)}\sum_{\myatop{t=1}{t\neq t_{\rm A}}}^{T(n)}U_t^{(n)}+\frac{U_{t_{\rm A}}^{(n,s)}}{T(n)},~~s\in\{0,1\}.
\end{align}
Let's denote by $\mu_{\rm U}(n)$ and $\sigma^2_{\rm U}(n)$ respectively the mean and
  variance of $U_{t_{\rm A}}^{(n,0)}$ and $U_t^{(n)}$, $t\neq t_{\rm A}$.
We define $g_n(x)$ as:
\begin{align}
\label{eq:g_n}g_n(x)=\frac{(x-(T(n)-1)\mu_{\rm U}(n))T(n)}{\sigma_{\rm U}(n)\sqrt{T(n)-1}}.
\end{align}
Thus, the re-scaled test statistic $L^{(n)}_s$ is expressed as follows:
\begin{align}
\label{eq:L0}L^{(n)}_0&=\frac{1}{\sqrt{T(n)-1}}\sum_{\myatop{t=1}{t\neq t_{\rm A}}}^{T(n)}\frac{U_t^{(n)}-\mu_{\rm U}(n)}{\sigma_{\rm U}(n)}+\frac{U_{t_{\rm A}}^{(n,0)}}{\sigma_{\rm U}(n)\sqrt{T(n)-1}}
\end{align}
when $H_0$ is true, and
\begin{align}
\label{eq:L1}L^{(n)}_1&=\frac{1}{\sqrt{T(n)-1}}\sum_{\myatop{t=1}{t\neq t_{\rm A}}}^{T(n)}\frac{U_t^{(n)}-\mu_{\rm U}(n)}{\sigma_{\rm U}(n)}+\frac{U_{t_{\rm A}}^{(n,1)}}{\sigma_{\rm U}(n)\sqrt{T(n)-1}}
\end{align}
when $H_1$ is true.
Provided $U_t^{(n)}$ satisfies the regularity conditions required by 
  the central limit theorem (CLT) for triangular arrays
  \cite[Theorem 27.2]{billingsley95measure},
  $\frac{1}{\sqrt{T(n)-1}}\sum_{\myatop{t=1}{t\neq t_{\rm A}}}^{T(n)}\frac{U_t^{(n)}-\mu_{\rm U}(n)}{\sigma_{\rm U}(n)}\xrightarrow{\mathcal{D}}Z$,
  where $Z\sim\mathcal{N}(0,1)$.
Thus, the weighted sum in \eqref{eq:L0} and \eqref{eq:L1} corresponds to 
  $S^{(n)}$ in Lemma \ref{lemma:LRT_bound}.
Now consider the term corresponding to slot $t_{\rm A}$,
  $\frac{U_{t_{\rm A}}^{(n,s)}}{\sigma_{\rm U}(n)\sqrt{T(n)-1}}$.
It effectively offsets $Z$'s mean away from zero and its distribution depends 
  on Alice's transmission state $s$.
Thus, depending on which hypothesis is true, it maps to either $V^{(n)}_0$ or 
  $V^{(n)}_1$ in Lemma \ref{lemma:LRT_bound}.
To prove achievability of covert communication, we show that there exists
  a coding scheme for Alice such that the random 
  variable describing the LRT statistic has the form given in $\eqref{eq:L0}$
  and $\eqref{eq:L1}$, with the terms in the sums satisfying the regularity
  conditions required by the CLT and
  the term corresponding to slot $t_{\rm A}$ converging to zero in probability.
This allows us to establish the covertness of Alice's transmission by applying
  Lemma \ref{lemma:LRT_bound}.
We prove reliability by extending the random coding arguments
  from \cite{bash12sqrtlawisit,bash13squarerootjsac}.

\subsection{Average Power Constraint}
We first show achievability under an average power constraint 
  $P_{\max}\in(0,\infty)$.
\begin{customtheorem}[Achievability]
\label{th:achievability}
Suppose Alice has a slotted AWGN channel to Bob with $T(n)=\omega(1)$ slots,
  each containing $n$ symbol periods, and that her transmitter is subject to
  the average power constraint $P_{\max}\in(0,\infty)$.
Then, provided that Alice and Bob share a sufficiently long secret,
  if Alice chooses to, she can transmit 
  $\mathcal{O}\left(\min\{\sqrt{n\log T(n)},n\}\right)$
  bits in a single slot while
  $\lim_{n\rightarrow\infty}\mathbb{P}_{\rm FA}+\mathbb{P}_{\rm MD}\geq1-\epsilon$ 
  and $\lim_{n\rightarrow\infty}\mathbb{P}_{\rm e}^{({\rm b})}\leq\delta$ for 
  arbitrary $\epsilon>0$ and $\delta>0$.
\end{customtheorem}

\begin{IEEEproof}
\textbf{Construction:} Alice secretly selects slot $t_{\rm A}$ uniformly at
  random out of the $T(n)$ slots.
Alice's channel encoder takes as input blocks of length $M$ bits and encodes
  them into codewords of length $n$ symbols.
We employ a random coding argument and independently generate $2^{M}$
  codewords $\{\mathbf{c}(W_k), k=1,2,\ldots,2^{M}\}$ from $\mathbb{R}^n$
  for messages $\{W_k\}_{k=1}^{2^{M}}$, each according
  to $p_{\mathbf{X}}(\mathbf{x})=\prod_{i=1}^{n}p_X(x_i)$, where
  $X\sim \mathcal{N}(0,P_{\rm f})$ and 
  symbol power $P_{\rm f}<\frac{\sigma_{\rm w}^2}{2}$ is defined later.
The codebook\footnote{Another way of viewing the construction is as a choice 
  of one of $T(n)$ codebooks, where the $i^{\text{th}}$ codebook has a block of
  non-zero symbols in the $i^{\text{th}}$ slot. 
Selection of the $t_{\rm A}$-th slot is equivalent to selection of 
  the $t_{\rm A}$-th codebook and the message is encoded by choosing a codeword
  from the selected codebook.}
is used only to send a single message and, along with
  $t_{\rm A}$, is the secret not revealed to Willie, though
  he knows how it is constructed, including the value of $P_{\rm f}$.

\textbf{Analysis (Willie):}
Denote by $\Upsilon_t=\sum_{Y_i\in \mathbf{Y}_{\rm w}(t)}Y_i^2$ the power in slot $t$.
Since Willie's channel from Alice is corrupted by AWGN with power $\sigma_{\rm w}^2$,
  the likelihood function of the observations $\mathbf{Y}_{\rm w}$ under $H_0$ is:
\begin{align}
\label{eq:likelihood_H0_s}f_0(\mathbf{Y}_{\rm w})&=\left(\frac{1}{2\pi\sigma_{\rm w}^2}\right)^{\frac{nT(n)}{2}}\exp\left[-\frac{1}{2\sigma_{\rm w}^2}\sum_{t=1}^{T(n)}\Upsilon_t\right].
\end{align}
Since Willie does not know which of the $T(n)$ slots Alice randomly
  selects for communication, or the codebook Alice and Bob use, but knows that
  Alice's signal is Gaussian, the likelihood function of the observations 
  $\mathbf{Y}_{\rm w}$ under $H_1$ is:
\begin{align}
f_1(\mathbf{Y}_{\rm w})&=\frac{1}{(2\pi\sigma^2_{\rm w})^{\frac{(T(n)-1)n}{2}}(2\pi(\sigma_{\rm w}^2+P_{\rm f}))^{\frac{n}{2}}T(n)}\nonumber\\
&\phantom{=}\times\label{eq:likelihood_H1_s}\sum_{t=1}^{T(n)}\exp\left[-\frac{\Upsilon_t}{2(\sigma_{\rm w}^2+P_{\rm f})}-\frac{\sum_{\myatop{r=1}{r\neq t}}^{T(n)}\Upsilon_r}{2\sigma_{\rm w}^2}\right].
\end{align}
The LRT statistic $\Lambda^{(n)}_s$ is the ratio between
  \eqref{eq:likelihood_H0_s} and \eqref{eq:likelihood_H1_s}.
Re-arranging terms yields:
  
\begin{align}
\label{eq:lrt_orig_s}
\Lambda^{(n)}_s&=\frac{1}{T(n)}\sum_{t=1}^{T(n)}\left(\frac{\sigma_{\rm w}^2}{\sigma_{\rm w}^2+P_{\rm f}}\right)^{\frac{n}{2}}\exp\left[{\frac{P_{\rm f}\Upsilon_t}{2\sigma_{\rm w}^2(\sigma_{\rm w}^2+P_{\rm f})}}\right].
\end{align}

When Alice does not transmit in the $i^{\text{th}}$ symbol period, 
  $Y_i\sim\mathcal{N}(0,\sigma_{\rm w}^2)$ since Willie observes AWGN;
  when Alice transmits, $Y_i\sim\mathcal{N}(0,\sigma_{\rm w}^2+P_{\rm f})$ by
  construction.
Let $\{X_t\}$, $X_t\sim\chi^2_n$, $t=1,\ldots,T(n)$ be a sequence of
  i.i.d.~chi-squared random 
  variables with $n$ degrees of freedom.
Then $\Upsilon_t=\sigma_{\rm w}^2X_t$ for all $t\in\{1,\ldots,T(n)\}$ under $H_0$ and
  $t\in\{1,\ldots,T(n)\}\setminus\{t_{\rm A}\}$ under $H_1$, while
  $\Upsilon_{t_{\rm A}}=(\sigma_{\rm w}^2+P_{\rm f})X_{t_{\rm A}}$ under $H_1$.
Let $U_t^{(n)}=\left(\frac{\sigma_{\rm w}^2}{\sigma_{\rm w}^2+P_{\rm f}}\right)^{\frac{n}{2}}\exp\left[\frac{P_{\rm f}X_t}{2(\sigma_{\rm w}^2+P_{\rm f})}\right]$ 
  for all $t\in\{1,\ldots,T(n)\}\setminus\{t_{\rm A}\}$, 
  $U_{t_{\rm A}}^{(n,0)}=\left(\frac{\sigma_{\rm w}^2}{\sigma_{\rm w}^2+P_{\rm f}}\right)^{\frac{n}{2}}\exp\left[\frac{P_{\rm f}X_t}{2(\sigma_{\rm w}^2+P_{\rm f})}\right]$,
  and
  $U_{t_{\rm A}}^{(n,1)}=\left(\frac{\sigma_{\rm w}^2}{\sigma_{\rm w}^2+P_{\rm f}}\right)^{\frac{n}{2}}\exp\left[\frac{P_{\rm f}X_t}{2\sigma_{\rm w}^2}\right]$.
Application of $g_n(x)$ in \eqref{eq:g_n} to \eqref{eq:lrt_orig_s} yields 
  the expression for $L^{(n)}_s$ in the form defined in \eqref{eq:L0} and 
  \eqref{eq:L1}.

Using the moment generating function (MGF)
  $\mathcal{M}_{\chi^2_n}(x)=(1-2x)^{-n/2}$ of a chi-squared random variable,
  we have:
\begin{align}
\mu_{\rm U}(n)&=\left(\frac{\sigma_{\rm w}^2}{\sigma_{\rm w}^2+P_{\rm f}}\right)^{\frac{n}{2}}\mathbb{E}\left[\exp\left(\frac{P_{\rm f}X_t}{2(\sigma_{\rm w}^2+P_{\rm f})}\right)\right]=1\\
\label{eq:sigma2U0}\sigma^2_{\rm U}(n)&=\left(\frac{\sigma_{\rm w}^2}{\sigma_{\rm w}^2+P_{\rm f}}\right)^{n}\mathbb{E}\left[\exp\left(\frac{P_{\rm f}X_t}{\sigma_{\rm w}^2+P_{\rm f}}\right)\right]-1=\left(\frac{\sigma_{\rm w}^4}{\sigma_{\rm w}^4-P_{\rm f}^2}\right)^{\frac{n}{2}}-1.
\end{align}
Since $P_{\rm f}<\frac{\sigma_{\rm w}^2}{2}$, $\mu_{\rm U}(n)$ and $\sigma^2_{\rm U}(n)$ 
  satisfy conditions for the CLT \cite[Theorem 27.2]{billingsley95measure},
\\
  $\frac{1}{\sqrt{T(n)-1}}\sum_{\myatop{t=1}{t\neq t_{\rm A}}}^{T(n)}\frac{U_t^{(n)}-\mu_{\rm U}(n)}{\sigma_{\rm U}(n)}\xrightarrow{\mathcal{D}}Z$, 
  where $Z\sim\mathcal{N}(0,1)$.
Also, when Alice does not transmit, by Chebyshev's inequality:
\begin{align}
\mathbb{P}\left(\left|\frac{U_{t_{\rm A}}^{(n,0)}}{\sigma_{\rm U}(n)\sqrt{T(n)-1}}\right|>\delta\right)&\leq \frac{\sigma^2_{\rm U}(n)}{(\delta\sigma_{\rm U}(n)\sqrt{T(n)-1}-1)^{2}}.\nonumber
\end{align}
Since $T(n)=\omega(1)$ and $P_{\rm f}<\frac{\sigma_{\rm w}^2}{2}$, 
  $\frac{U_{t_{\rm A}}^{(n,0)}}{\sigma_{\rm U}(n)\sqrt{T(n)-1}}\xrightarrow{\mathcal{P}}0$
  as $n\rightarrow\infty$.

When Alice transmits,
\begin{align}
\label{eq:EUtA}\expected{U_{t_{\rm A}}^{(n,1)}}&=\left(\frac{\sigma_{\rm w}^2}{\sigma_{\rm w}^2+P_{\rm f}}\right)^{\frac{n}{2}}\mathbb{E}\left[\exp\left(\frac{P_{\rm f}X_t}{2\sigma_{\rm w}^2}\right)\right]=\left(\frac{\sigma_{\rm w}^4}{\sigma_{\rm w}^4-P_{\rm f}^2}\right)^{\frac{n}{2}}\\
\label{eq:EUtA2}\expected{\left(U_{t_{\rm A}}^{(n,1)}\right)^2}&=\left(\frac{\sigma_{\rm w}^2}{\sigma_{\rm w}^2+P_{\rm f}}\right)^{n}\mathbb{E}\left[\exp\left(\frac{P_{\rm f}X_t}{\sigma_{\rm w}^2}\right)\right]=\left(\frac{\sigma_{\rm w}^2}{\sigma_{\rm w}^2+P_{\rm f}}\right)^{n}\left(\frac{\sigma_{\rm w}^2}{\sigma_{\rm w}^2-2P_{\rm f}}\right)^{\frac{n}{2}}.
\end{align}
To complete the analysis of Willie's detector, we must show
  $\frac{U_{t_{\rm A}}^{(n,1)}}{\sigma_{\rm U}(n)\sqrt{T(n)-1}}\xrightarrow{\mathcal{P}}0$
  as $n\rightarrow\infty$.
By Chebyshev's inequality:
\begin{align}
\mathbb{P}\left(\left|\frac{U_{t_{\rm A}}^{(n,1)}}{\sigma_{\rm U}(n)\sqrt{T(n)-1}}\right|>\delta\right) & \leq \frac{\Var\left[U_{t_{\rm A}}^{(n,1)}\right]}{\left(\delta\sigma_{\rm U}(n)\sqrt{T(n)-1}-\expected{U_{t_{\rm A}}^{(n,1)}}\right)^{2}}\nonumber\\
\label{eq:cheb1}&=\left(\frac{\delta\sigma_{\rm U}(n)\sqrt{T(n)-1}}{\sqrt{\Var\left[U_{t_{\rm A}}^{(n,1)}\right]}}-\frac{\expected{U_{t_{\rm A}}^{(n,1)}}}{\sqrt{\Var\left[U_{t_{\rm A}}^{(n,1)}\right]}}\right)^{-2}.
\end{align}
Thus, it is sufficient to show that, as long as $n\rightarrow\infty$ and 
  $T(n)=\omega(1)$, 
\begin{align}
\label{eq:cond1}\expected{U_{t_{\rm A}}^{(n,1)}}\bigg/\sqrt{\Var\left[U_{t_{\rm A}}^{(n,1)}\right]}&\rightarrow 0, \text{~and}\\
\label{eq:cond2}\sigma^2_{\rm U}(n)(T(n)-1)\bigg/\Var\left[U_{t_{\rm A}}^{(n,1)}\right]&\rightarrow\infty.
\end{align}
We use \eqref{eq:sigma2U0}, \eqref{eq:EUtA} and \eqref{eq:EUtA2} 
  to obtain:
\begin{align}
\label{eq:rat0}\frac{\expected{U_{t_{\rm A}}^{(n,1)}}}{\sqrt{\Var\left[U_{t_{\rm A}}^{(n,1)}\right]}}&=\frac{1}{\sqrt{\left(1+\frac{P_{\rm f}^2}{\sigma_{\rm w}^4(1-2P_{\rm f}/\sigma_{\rm w}^2)}\right)^{\frac{n}{2}}-1}}\\
\label{eq:erat0}\frac{\sigma^2_{\rm U}(n)}{\Var\left[U_{t_{\rm A}}^{(n,1)}\right]}&\geq \frac{\sigma^2_{\rm U}(n)}{\expected{\left(U_{t_{\rm A}}^{(n,1)}\right)^2}}\\
\label{eq:erat1}&=\left(1-\frac{2P_{\rm f}^2}{\sigma_{\rm w}^4(1-P_{\rm f}/\sigma_{\rm w}^2)}\right)^{\frac{n}{2}}-\left(\frac{\sigma_{\rm w}^2}{\sigma_{\rm w}^2+P_{\rm f}}\right)^{-n}\left(\frac{\sigma_{\rm w}^2}{\sigma_{\rm w}^2-2P_{\rm f}}\right)^{-\frac{n}{2}}\\
\label{eq:erat2}&=\left(1-\frac{2P_{\rm f}^2}{\sigma_{\rm w}^4(1-P_{\rm f}/\sigma_{\rm w}^2)}\right)^{\frac{n}{2}}-o(1),
\end{align}
where \eqref{eq:erat2} follows since $\sigma_{\rm w}^2>0$ and 
  $P_{\rm f}$ satisfies $0\leq P_{\rm f}<\sigma_{\rm w}^2/2$ by construction.

Now consider two scaling regimes for $T(n)$:
\begin{itemize}
\item $T(n)=o(e^n)$:
  set $P_{\rm f}=\frac{c_{\rm P}^{(\mathrm{S})}\sigma_{\rm w}^2\sqrt{\log T(n)}}{\sqrt{n}}$
  with $c_{\rm P}^{(\mathrm{S})}>0$ a constant determined later.
Taylor series expansion of $\log(1+x)$ at $x=0$ yields:
\begin{align}
\left(1+\frac{P_{\rm f}^2}{\sigma_{\rm w}^4(1-2P_{\rm f}/\sigma_{\rm w}^2)}\right)^{\frac{n}{2}}&=e^{\frac{n}{2}\log\left(1+\frac{P_{\rm f}^2}{\sigma_{\rm w}^4(1-2P_{\rm f}/\sigma_{\rm w}^2)}\right)}=e^{\frac{c_{\rm P}^{(\mathrm{S})}}{2}\log T(n)-o(\log T(n))},\nonumber
\end{align}
  implying that \eqref{eq:rat0} converges to zero since $T(n)=\omega(1)$.
Thus, \eqref{eq:cond1} holds.
Furthermore, Taylor series expansion of $\log(1-x)$ at $x=0$ shows:
\begin{align}
\left(1-\frac{2P_{\rm f}^2}{\sigma_{\rm w}^4(1-P_{\rm f}/\sigma_{\rm w}^2)}\right)^{\frac{n}{2}}&=e^{\frac{n}{2}\log\left(1-\frac{2P_{\rm f}^2}{\sigma_{\rm w}^4(1-P_{\rm f}/\sigma_{\rm w}^2)}\right)}=e^{-c_{\rm P}^{(\mathrm{S})}\log T(n)-o(\log T(n))}\nonumber,
\end{align}
Thus, there exists $c_{\rm P}^{(\mathrm{S})}\in(0,1)$ such that \eqref{eq:cond2}
  holds.

\item $T(n)=\Omega(e^n)$: set
  $P_{\rm f}=c_{\rm P}^{(\mathrm{L})}\sigma_{\rm w}^2/2$ with
  $c_{\rm P}^{(\mathrm{L})}\in(0,1)$ a constant.
Then, clearly, \eqref{eq:cond1} holds since \eqref{eq:rat0} converges to zero,
  and \eqref{eq:cond2} holds for an appropriately chosen 
  $c_{\rm P}^{(\mathrm{L})}$.
\end{itemize}

Therefore, setting
  $P_{\rm f}=\sigma_{\rm w}^2\min\{\frac{c_{\rm P}^{(\mathrm{S})}\sqrt{\log T(n)}}{\sqrt{n}},\frac{c_{\rm P}^{(\mathrm{L})}}{2}\}$
  ensures convergence of the RHS of \eqref{eq:cheb1} to zero for any scaling 
  of $T(n)$, and, by Lemma \ref{lemma:LRT_bound},
  ensures $\mathbb{P}_{\rm FA}+\mathbb{P}_{\rm MD}\geq1-\epsilon$ 
  for any $\epsilon>0$.

\textbf{Analysis (Bob):} 
Let Bob employ the maximum likelihood (ML) decoder (i.e., minimum distance
  decoder).
If Bob knows the value of $t_{\rm A}$, 
  Alice can reliably (i.e., with Bob's decoding error probability, averaged over
  all the codebooks, decaying to zero as $n\rightarrow\infty$) transmit 
  $M=\frac{n\gamma}{2}\log_2\left(1+\frac{\sigma_{\rm w}^2}{2\sigma_{\rm b}^2}\min\left\{\frac{c_{\rm P}^{(\mathrm{S})}\sqrt{\log T(n)}}{\sqrt{n}},\frac{c_{\rm P}^{(\mathrm{L})}}{2}\right\}\right)$
  covert bits, where $\gamma\in(0,1)$ is a constant
  \cite{bash12sqrtlawisit,bash13squarerootjsac}.
However, knowledge of $t_{\rm A}$ is unnecessary for Bob if 
  $T(n)<2^{c_{\rm T}n}$, where $c_{\rm T}>0$ is a constant, as we show next.
Let's augment Alice and Bob's Gaussian codebook with the origin 
  $\mathbf{c}(0)=\{0,\ldots,0\}$ 
  (indicating ``no transmission'') and have Bob attempt to decode each of the
  $T(n)$ slots.
The squared distance between a codeword $\mathbf{c}(W_k)$ and $\mathbf{c}(0)$ is
  $P_{\rm f}X$,
  where $X\sim\chi^2_n$.
Repeating the analysis of Bob's detection error probability 
  from~\cite{bash12sqrtlawisit,bash13squarerootjsac} using the 
  distance between $\mathbf{c}(W_k)$ and $\mathbf{c}(0)$ instead of 
  $\mathbf{c}(W_i)$ yields a looser upper bound on the
  probability of the decoding error in each slot.
By the union bound over all $T(n)$ slots, the overall probability of error is
$\mathbb{P}_{\rm e}^{({\rm b})}\leq T(n)2^{M-\frac{n}{2}\log_2\left(1+\frac{P_{\rm f}}{4\sigma_{\rm b}^2}\right)}$.
If $T(n)=o(e^n)$, then clearly Bob's decoding error probability decays to zero
  if Alice attempts to transmit 
  $M=\frac{n\gamma}{2}\log_2\left(1+\frac{c_{\rm P}^{(\mathrm{S})}\sigma_{\rm w}^2\sqrt{\log T(n)}}{4\sigma_{\rm b}^2\sqrt{n}}\right)$
  bits in a randomly selected $n$-symbol slot $t_{\rm A}$.
If $T(n)=\Omega(e^n)$, then, $P_{\rm f}=\frac{c_{\rm P}^{(\mathrm{L})}\sigma_{\rm w}^2}{2}$, and 
  $T(n)<2^{c_{\rm T}n}$ where
  $c_{\rm T}=\frac{1-\gamma}{2}\log_2\left(1+\frac{c_{\rm P}^{(\mathrm{L})}\sigma_{\rm w}^2}{8\sigma_{\rm b}^2}\right)$
  ensures that Bob's decoding error probability decays to zero
  if Alice attempts to transmit 
  $M=\frac{n\gamma}{2}\log_2\left(1+\frac{c_{\rm P}^{(\mathrm{L})}\sigma_{\rm w}^2}{8\sigma_{\rm b}^2}\right)$
  bits in a randomly selected $n$-symbol slot $t_{\rm A}$.
Therefore, $\mathcal{O}(\min\{\sqrt{n\log T(n)},n\})$ covert bits can be 
  transmitted reliably using slot $t_{\rm A}$.
\end{IEEEproof}

\noindent \emph{Remark:} The logarithm of 
  \eqref{eq:lrt_orig_s} is the log-likelihood ratio:
\begin{align*}
\log \Lambda^{(n)}_s&=-\log T(n)+u(n)\log\sum_{t=1}^{T(n)}\exp\left[v(n)\Upsilon_t\right]\\
&=-\log T(n)+u(n)\operatorname{LogSumExp}(\{v(n)\Upsilon_t\}_{t=1}^{T(n)}),
\end{align*}
where $u(n)\equiv\left(\frac{\sigma_{\rm w}^2}{\sigma_{\rm w}^2+P_{\rm f}}\right)^{\frac{n}{2}}$ 
  and $v(n)\equiv\frac{P_{\rm f}}{2\sigma_{\rm w}^2(\sigma_{\rm w}^2+P_{\rm f})}$.
Since $\operatorname{LogSumExp}(\{x_i\})$ is an analytic 
  approximation of $\max(\{x_i\})$ \cite[Ch.~3.1.5]{boyd04convexopt},
  Willie's (approximate) sufficient statistic is the maximum slot power
  $\Upsilon_{\max}=\max_{t\in\{1,\ldots,T(n)\}}\Upsilon_t$.
While this motivates the design of Willie's detector in the converse proof, 
  in the achievability proofs we analyze the exact LRT.

\subsection{Peak Power Constraint}
Unfortunately, representing real-valued codewords requires unbounded storage,
  which means that the length of the secret pre-shared by Alice and Bob is
  infinite.
To address this, we consider a finite alphabet, which also satisfies 
  a peak power constraint $P_{\max}\in(0,\infty)$ on the transmitter.
The remark in~\cite[Sec.~III]{bash13squarerootjsac} allows both improvement 
  of Bob's decoding performance and reduction of the size of the pre-shared 
  secret to $\mathcal{O}(n)$ bits (provided $T(n)<2^{c_{\rm T}n}$).
Approaches reported in \cite{bash13squarerootjsac,bloch15covert,wang15covert}
  may reduce the pre-shared secret to $\mathcal{O}(\sqrt{n}\log n)$ bits, and
  even possibly to $\mathcal{O}(\sqrt{n})$ bits.

\begin{customtheorem}[Achievability under peak power constraint]
\label{th:achievability_peak_power}
Suppose Alice has a slotted AWGN channel to Bob with $T(n)=\omega(1)$ slots,
  each containing $n$ symbol periods, and that her transmitter is subject to
  the peak power constraint $P_{\max}\in(0,\infty)$.
Then, provided that Alice and Bob share a sufficiently long secret,
  if Alice chooses to, she can transmit 
  $\mathcal{O}\left(\min\{\sqrt{n\log T(n)},n\}\right)$
  bits in a single slot while
  $\lim_{n\rightarrow\infty}\mathbb{P}_{\rm FA}+\mathbb{P}_{\rm MD}\geq1-\epsilon$
  and $\lim_{n\rightarrow\infty}\mathbb{P}_{\rm e}^{({\rm b})}\leq\delta$ for 
  any $\epsilon>0$ and $\delta>0$.
\end{customtheorem}

\begin{IEEEproof}
\textbf{Construction:} Alice secretly selects slot $t_{\rm A}$ uniformly at
  random out of the $T(n)$ slots in which to communicate.
She encodes the input in blocks of length $M$ bits into codewords of length 
  $n$ symbols with the symbols drawn
  from alphabet $\{-a,a\}$, where $a$ satisfies the peak power constraint 
  $a^2<P_{\max}$ and is defined later. 
Alice independently generates $2^{M}$ codewords 
  $\{\mathbf{c}(W_k),k=1,2,\ldots,2^{M}\}$ for messages
  $\{W_k\}$ from $\{-a,a\}^{n}$ according to 
  $p_{\mathbf{X}}(\mathbf{x})=\prod_{i=1}^{n}p_X(x_i)$, where
  $p_X(-a)=p_X(a)=\frac{1}{2}$.
As in the proof of Theorem \ref{th:achievability},
  this single-use codebook is not revealed to Willie, though he knows how it is
  constructed, including the value of $a$.
While in this proof the entire codebook is secretly shared between Alice and 
  Bob, the amount of shared secret information can be reduced using the remark
  in~\cite[Sec.~III]{bash13squarerootjsac}.

\textbf{Analysis (Willie):}
Since the model for the AWGN channel from Alice to Willie is the same as in 
  Theorem \ref{th:achievability}, the likelihood function of 
  the observations $\mathbf{Y}_{\rm w}$ under $H_0$ is given by 
  \eqref{eq:likelihood_H0_s}.
Since Willie does not know which of the $T(n)$ slots Alice randomly
  selects for communication, or the codebook Alice and Bob use, but knows how
  the codebook is constructed, the likelihood function of the observations 
  $\mathbf{Y}_{\rm w}$ under $H_1$ is:
\begin{align}
\label{eq:likelihood_H1_s_pp}f_1(\mathbf{Y}_{\rm w})&=\frac{1}{(2\pi\sigma_{\rm w}^2)^{\frac{nT(n)}{2}}T(n)}\sum_{t=1}^{T(n)}\frac{1}{2^n}\sum_{\mathbf{b}\in\{-1,1\}^n}e^{-\frac{\sum_{\myatop{r=1}{r\neq t}}^{T(n)}\Upsilon_r+\sum_{Y_i\in\mathbf{Y}_{\rm w}(t)}(Y_i-ab_i)^2}{2\sigma_{\rm w}^2}},
\end{align}
  where $\Upsilon_t=\sum_{Y_i\in \mathbf{Y}_{\rm w}(t)}Y_i^2$ denotes the power in 
  slot $t$.
The LRT statistic $\Lambda^{(n)}_s$ is the ratio between
  \eqref{eq:likelihood_H0_s} and \eqref{eq:likelihood_H1_s_pp}.
Re-arranging terms yields:
\begin{align}
\label{eq:lrt_orig_s_pp}
\Lambda^{(n)}_s&=\frac{1}{T(n)}\sum_{t=1}^{T(n)}\frac{\exp\left[{-\frac{na^2}{2\sigma_{\rm w}^2}}\right]}{2^n}\sum_{\mathbf{b}\in\{-1,1\}^n}e^{{\frac{a}{\sigma_{\rm w}^2}\sum_{Y_i\in\mathbf{Y}_{\rm w}(t)}Y_ib_i}}.
\end{align}
Let $U^{(n)}_t=\frac{\exp\left[{-\frac{na^2}{2\sigma_{\rm w}^2}}\right]}{2^n}\sum_{\mathbf{b}\in\{-1,1\}^n}\exp\left[{\frac{a}{\sigma_{\rm w}^2}\sum_{Y_i\in\mathbf{Y}_{\rm w}(t)}Y_ib_i}\right]$.
Application of $g_n(x)$ in \eqref{eq:g_n} to \eqref{eq:lrt_orig_s_pp} yields 
  the expression for $L^{(n)}_s$ in the form defined in \eqref{eq:L0} and 
  \eqref{eq:L1}.

When Alice does not transmit,
$\exp\left[{\frac{a}{\sigma_{\rm w}^2}\sum_{Y_i\in\mathbf{Y}_{\rm w}(t)}Y_ib_i}\right]\sim\log\mathcal{N}\left(0,\frac{na^2}{\sigma_{\rm w}^2}\right)$,
  where $\log\mathcal{N}(\mu,\sigma^2)$ denotes the log-normal distribution with
  location $\mu$ and scale $\sigma^2$.
Thus,
\begin{align}
\mu_{\rm U}(n)&=\frac{\exp\left[{-\frac{na^2}{2\sigma_{\rm w}^2}}\right]}{2^n}\sum_{\mathbf{b}\in\{-1,1\}^n}\expected{e^{{\frac{a}{\sigma_{\rm w}^2}\sum_{Y_i\in\mathbf{Y}_{\rm w}(t)}Y_ib_i}}}=1.\nonumber
\end{align}
To obtain $\sigma_{\rm U}^2(n)$, we calculate the second moment of $U^{(n,0)}_{t_{\rm A}}$ 
  and $U^{(n)}_t$, $t\neq t_{\rm A}$:
\begin{align}
\label{eq:calc_su1}\expected{\left(U^{(n)}_t\right)^2}&=\frac{e^{{-\frac{na^2}{\sigma_{\rm w}^2}}}}{2^{2n}}\sum_{\mathbf{b},\mathbf{d}\in\{-1,1\}^n}\expected{e^{{\frac{a}{\sigma_{\rm w}^2}\sum_{Y_i\in\mathbf{Y}_{\rm w}(t)}Y_i(b_i+d_i)}}}=\frac{1}{2^{2n}}\sum_{\mathbf{b},\mathbf{d}\in\{-1,1\}^n}e^{\frac{a^2}{\sigma_{\rm w}^2}\sum_{i=1}^nb_id_i}\\
\label{eq:calc_su2}&=\frac{1}{2^{2n}}\sum_{\mathbf{b},\mathbf{z}\in\{-1,1\}^n}e^{\frac{a^2}{\sigma_{\rm w}^2}\sum_{i=1}^nz_i}\\
\label{eq:calc_su3}&=\cosh^n\left(\frac{a^2}{\sigma_{\rm w}^2}\right),
\end{align}
where \eqref{eq:calc_su1} follows from
  $\exp\left[{\frac{a}{\sigma_{\rm w}^2}\sum_{Y_i\in\mathbf{Y}_{\rm w}(t)}Y_i(b_i+d_i)}\right]\sim\log\mathcal{N}\left(0,\frac{a^2}{\sigma_{\rm w}^2}\sum_{i=1}^n(b_i+d_i)^2\right)$; 
  \eqref{eq:calc_su2} is since for a given
  $\mathbf{b}\in\{-1,1\}^n$ and any $\mathbf{d}\in\{-1,1\}^n$,
  $\mathbf{z}=[b_1d_1,b_2d_2,\ldots,b_nd_n]\in\{-1,1\}^n$ is unique;
  and \eqref{eq:calc_su3} follows from Appendix \ref{app:derive_cosh}.
Thus, $\frac{1}{\sqrt{T(n)-1}}\sum_{\myatop{t=1}{t\neq t_{\rm A}}}^{T(n)}\frac{U_t^{(n)}-\mu_{\rm U}(n)}{\sigma_{\rm U}(n)}\xrightarrow{\mathcal{D}}Z$, 
  $Z\sim\mathcal{N}(0,1)$.
Also, when Alice does not transmit, by Chebyshev's inequality:
\begin{align}
\mathbb{P}\left(\left|\frac{U_{t_{\rm A}}^{(n,0)}}{\sigma_{\rm U}(n)\sqrt{T(n)-1}}\right|>\delta\right)&\leq \frac{\sigma^2_{\rm U}(n)}{(\delta\sigma_{\rm U}(n)\sqrt{T(n)-1}-1)^{2}}\nonumber.
\end{align}
Since $T(n)=\omega(1)$, 
  $\frac{U_{t_{\rm A}}^{(n,0)}}{\sigma_{\rm U}(n)\sqrt{T(n)-1}}\xrightarrow{\mathcal{P}}0$
  as $n\rightarrow\infty$.

When Alice transmits, by construction, Willie observes 
  $y_{(t_{\rm A}-1)n+i}\sim\mathcal{N}(ac_i,\sigma_{\rm w}^2)$, $i=1,\ldots,n$, where 
  $\mathbf{c}=[c_1,\ldots,c_n]$ is drawn equiprobably from $\{-1,1\}^n$.
Thus,
\begin{align}
\expected{U_{t_{\rm A}}^{(n,1)}}&=\frac{e^{-\frac{na^2}{2\sigma_{\rm w}^2}}}{2^{2n}}\sum_{\mathbf{c},\mathbf{b}\in\{-1,1\}^n}\expected{e^{\frac{a}{\sigma_{\rm w}^2}\sum_{Y_i\in\mathbf{Y}_{\rm w}(t)}Y_ib_i}}\nonumber\\
\label{eq:calc_muta}&=\frac{1}{2^{2n}}\sum_{\mathbf{c},\mathbf{b}\in\{-1,1\}^n}e^{\frac{a^2}{\sigma_{\rm w}^2}\sum_{i=1}^nb_ic_i}=\cosh^n\left(\frac{a^2}{\sigma_{\rm w}^2}\right),
\end{align}
where \eqref{eq:calc_muta} follows from
  $\exp\left[{\frac{a}{\sigma_{\rm w}^2}\sum_{Y_i\in\mathbf{Y}_{\rm w}(t)}Y_ib_i}\right]\sim\log\mathcal{N}\left(\frac{a^2}{\sigma_{\rm w}^2}\sum_{i=1}^nc_ib_i,\frac{na^2}{\sigma_{\rm w}^2}\right)$,
  the argument for \eqref{eq:calc_su2} above, and Appendix 
  \ref{app:derive_cosh}.
By the definition of variance and the law of total expectation,
  $\Var[U_{t_{\rm A}}^{(n,1)}]\leq\frac{1}{2^n}\sum_{\mathbf{c}\in\{-1,1\}^n}\expected{\left(U_{t_{\rm A}}^{(n,1)}\big|\mathbf{c}\text{~sent}\right)^2}$, where 
\begin{align}
\expected{\left(U_{t_{\rm A}}^{(n,1)}\big|\mathbf{c}\text{~sent}\right)^2}&=\frac{e^{-\frac{na^2}{\sigma_{\rm w}^2}}}{2^{2n}}\sum_{\mathbf{b},\mathbf{d}\in\{-1,1\}^n}\expected{e^{\frac{a}{\sigma_{\rm w}^2}\sum_{Y_i\in\mathbf{Y}_{\rm w}(t)}Y_i(b_i+d_i)}}\nonumber\\
&=\frac{e^{-\frac{na^2}{\sigma_{\rm w}^2}}}{2^{2n}}\sum_{\mathbf{b},\mathbf{d}\in\{-1,1\}^n}e^{\frac{a^2}{\sigma_{\rm w}^2}\sum_{i=1}^nc_i(b_i+d_i)+\frac{(b_i+d_i)^2}{2}}\nonumber\\
\label{eq:calc_suta}&=\frac{1}{2^{2n}}\sum_{\mathbf{b},\mathbf{d}\in\{-1,1\}^n}e^{\frac{a^2}{\sigma_{\rm w}^2}\sum_{i=1}^nc_ib_i+(c_i+b_i)d_i}\leq\cosh^n\left(\frac{a^2}{\sigma_{\rm w}^2}\right)\cosh^n\left(\frac{2a^2}{\sigma_{\rm w}^2}\right),
\end{align}
with \eqref{eq:calc_suta} following from $c_i+b_i\leq2$ and arguments for 
  \eqref{eq:calc_muta} above.
By Chebyshev's inequality:
\begin{align}
\label{eq:UT_bound1_pp}\mathbb{P}\left(\left|\frac{U_{t_{\rm A}}^{(n,1)}}{\sigma_{\rm U}(n)\sqrt{T(n)-1}}\right|>\delta\right)&\leq \frac{\cosh^n\left(\frac{a^2}{\sigma_{\rm w}^2}\right)\cosh^n\left(\frac{2a^2}{\sigma_{\rm w}^2}\right)}{\left(\delta\sigma_{\rm U}(n)\sqrt{T(n)-1}-\expected{U_{t_{\rm A}}^{(n,1)}}\right)^{2}}.
\end{align}

Dividing both numerator and denominator of \eqref{eq:UT_bound1_pp} by
  $\cosh^n\left(\frac{a^2}{\sigma_{\rm w}^2}\right)\cosh^n\left(\frac{2a^2}{\sigma_{\rm w}^2}\right)$,
  we note that $\frac{\expected{U_{t_{\rm A}}^{(n,1)}}}{\cosh^{\frac{n}{2}}\left(\frac{a^2}{\sigma_{\rm w}^2}\right)\cosh^{\frac{n}{2}}\left(\frac{2a^2}{\sigma_{\rm w}^2}\right)}=\left(\frac{\cosh(a^2/\sigma_{\rm w}^2)}{\cosh(2a^2/\sigma_{\rm w}^2)}\right)^{\frac{n}{2}}\leq 1$,
  as $\frac{\cosh(x)}{\cosh(2x)}\leq 1$ for $x\in\mathbb{R}$.
Also, $\frac{\sigma_{\rm U}^2(n)}{\cosh^{n}\left(\frac{a^2}{\sigma_{\rm w}^2}\right)\cosh^{n}\left(\frac{2a^2}{\sigma_{\rm w}^2}\right)}=\cosh^{-n}\left(\frac{2a^2}{\sigma_{\rm w}^2}\right)$.
When $T(n)=o(e^n)$, setting $a^2=\frac{c_{\rm P}^{(\mathrm{S})}\sigma_{\rm w}^2\sqrt{\log T(n)}}{\sqrt{2n}}$
  for a constant $c_{\rm P}^{(\mathrm{S})}\in(0,1)$ ensures 
  $\frac{U_{t_{\rm A}}^{(n,1)}}{\sigma_{\rm U}(n)\sqrt{T(n)-1}}\xrightarrow{\mathcal{P}}0$
  as $n\rightarrow\infty$.
Convergence follows from noting that 
$\cosh^{-n}\left(\frac{2a^2}{\sigma_{\rm w}^2}\right)=\exp\left[-n\log\cosh\left(\frac{2a^2}{\sigma_{\rm w}^2}\right)\right]\geq\exp\left[-\frac{2na^4}{\sigma_{\rm w}^4}\right]$.
When $T(n)=\Omega(e^n)$, convergence is obtained by setting
  $a^2=\frac{c_{\rm P}^{(\mathrm{L})}\sigma_{\rm w}^2}{2}$.
Therefore, by Lemma \ref{lemma:LRT_bound}, setting
  $a^2=\sigma_{\rm w}^2\min\left\{\frac{c_{\rm P}^{(\mathrm{S})}\sqrt{\log T(n)}}{\sqrt{2n}},\frac{c_{\rm P}^{(\mathrm{L})}}{2}\right\}$
  ensures $\mathbb{P}_{\rm FA}+\mathbb{P}_{\rm MD}\geq1-\epsilon$
  for any $\epsilon>0$.

  \textbf{Analysis (Bob):}
Suppose Alice transmits $\mathbf{c}(W_k)$.
As in the proof of Theorem \ref{th:achievability}, let Bob employ the ML decoder
  that suffers an error event $E_{k\rightarrow i}$ when the received vector 
  $\mathbf{Y}_{\rm b}$ is closer in Euclidean distance to codeword $\mathbf{c}(W_i)$, 
  $i\neq k$.
Knowledge of $t_{\rm A}$ ensures reliable decoding by the application of
  Appendix \ref{app:corr}: Bob's error probability,
  averaged over all the codebooks, decays to zero as
  $n\rightarrow\infty$ for a covert transmission containing
  $M=n\gamma\left(1-\log_2\left[1+\exp\left(-\frac{\sigma_{\rm w}^2}{2\sigma_{\rm b}^2}\min\left\{\frac{c_{\rm P}^{(\mathrm{S})}\sqrt{\log T(n)}}{\sqrt{2n}},\frac{c_{\rm P}^{(\mathrm{L})}}{2}\right\}\right)\right]\right)$
  bits, where $\gamma\in(0,1)$ is a constant.

However, as in Theorem \ref{th:achievability}, knowledge of $t_{\rm A}$ is 
  unnecessary for Bob if $T(n)<2^{c_{\rm T}n}$ where $c_{\rm T}$ is a constant.
Again, let's augment Alice and Bob's codebook with the origin 
  $\mathbf{c}(0)=\{0,\ldots,0\}$ 
  (indicating ``no transmission'') and have Bob attempt to decode each of the
  $T(n)$ slots.
Denoting the decoding error probability in slot $t$ by 
  $\mathbb{P}_{\rm e}^{({\rm b})}(t)$, we employ the union bound over all slots to 
  upper-bound the overall decoding error probability:
\begin{align}
\label{eq:p_e_b_1}\mathbb{P}_{\rm e}^{({\rm b})}&\leq\sum_{t=1}^{T(n)}\mathbb{P}_{\rm e}^{({\rm b})}(t).
\end{align}
The decoding error probability for one of the $T(n)-1$ slots that Alice
  does not use is the probability that the received vector is closer to
  some codeword than the origin $\mathbf{c}(0)$:
\begin{align}
\mathbb{P}_{\rm e}^{({\rm b})}(t)&=\mathbb{P}\left(\cup_{i=1}^{2^M}E_{0\rightarrow i}\right)\leq \sum_{i=1}^{2^M}\mathbb{P}(E_{0\rightarrow i}),t\neq t_{\rm A},\label{eq:p_e_b_blank}
\end{align}
where the inequality is the union bound.
Since the Euclidean distance between each codeword and $\mathbf{c}(0)$ is 
  $\sqrt{n}a$, by \cite[Eq.~(3.44)]{madhow08digicom}:
\begin{align}
\mathbb{P}(E_{0\rightarrow i})&=Q\left(\frac{\sqrt{n}a}{2\sigma_{\rm b}}\right)\leq\frac{1}{2}\exp\left(-\frac{na^2}{8\sigma_{\rm b}^2}\right),\label{eq:p_e_blank_to_codeword}
\end{align}
where $Q(x)=\frac{1}{\sqrt{2\pi}}\int_{x}^{\infty}e^{-t^2/2}dt$ and 
  the inequality is because of the upper bound 
  $Q(x)\leq\frac{1}{2}e^{-x^2/2}$ \cite[Eq.~(5)]{chiani03qfunction}.
Substituting \eqref{eq:p_e_blank_to_codeword} into \eqref{eq:p_e_b_blank} 
  yields:
\begin{align}
\label{eq:p_e_b_blank_ub}\mathbb{P}_{\rm e}^{({\rm b})}(t)&\leq 2^{M-\frac{na^2\log_2e}{8\sigma_{\rm b}^2}},t\neq t_{\rm A}.
\end{align}

To upper-bound the decoding error probability for the slot that Alice uses to
  transmit, we combine the bounds in \eqref{eq:p_e_b_blank_ub} and
  \eqref{eq:p_e_b_ub} as follows:
\begin{align}
\label{eq:p_e_b_slot_ub}\mathbb{P}_{\rm e}^{({\rm b})}(t_{\rm A})\leq 2^{M-n\min\left\{1-\log_2\left[1+\exp\left(-\frac{a^2}{2\sigma_{\rm b}^2}\right)\right],\frac{a^2\log_2e}{8\sigma_{\rm b}^2}\right\}}.
\end{align}
Combining \eqref{eq:p_e_b_1}, \eqref{eq:p_e_b_blank_ub}, and 
  \eqref{eq:p_e_b_slot_ub} yields:
\begin{align}
\mathbb{P}_{\rm e}^{({\rm b})}&\leq (T(n)-1)2^{M-\frac{na^2\log_2e}{8\sigma_{\rm b}^2}}+2^{M-n\min\left\{1-\log_2\left[1+\exp\left(-\frac{a^2}{2\sigma_{\rm b}^2}\right)\right],\frac{a^2\log_2e}{8\sigma_{\rm b}^2}\right\}}.
\end{align}

If $T(n)=o(e^n)$, then Alice sets 
  $a^2=\frac{c_{\rm P}^{(\mathrm{S})}\sigma_{\rm w}^2\sqrt{\log T(n)}}{\sqrt{2n}}$.
Thus, for large enough $n$, 
  $\frac{a^2\log_2e}{8\sigma_{\rm b}^2}<1-\log_2\left[1+\exp\left(-\frac{a^2}{2\sigma_{\rm b}^2}\right)\right]$
  and Bob's decoding error probability decays to zero
  if Alice attempts to transmit 
  $M=\frac{\gamma c_{\rm P}^{(\mathrm{S})}\sigma_{\rm w}^2\sqrt{n\log T(n)}\log_2e}{8\sqrt{2}\sigma_{\rm b}^2}$
  bits in a randomly selected $n$-symbol slot $t_{\rm A}$, where $\gamma\in(0,1)$.
If $T(n)=\Omega(e^n)$, then, $a^2=\frac{c_{\rm P}^{(\mathrm{L})}\sigma_{\rm w}^2}{2}$, and 
  $T(n)<2^{c_{\rm T}n}$ where
  $c_{\rm T}=\frac{(1-\gamma)c_{\rm P}^{(\mathrm{L})}\sigma_{\rm w}^2\log_2e}{16\sigma_{\rm b}^2}$
  ensures that Bob's decoding error probability decays to zero
  if Alice attempts to transmit 
  $M=n\gamma\min\left\{1-\log_2\left[1+\exp\left(-\frac{\sigma_{\rm w}^2}{4\sigma_{\rm b}^2}\right)\right],\frac{c_{\rm P}^{(\mathrm{L})}\sigma_{\rm w}^2\log_2e}{16\sigma_{\rm b}^2}\right\}$
  bits in slot $t_{\rm A}$.
\end{IEEEproof}

\section{Converse}
\label{sec:converse}
In this section we show that Alice cannot transmit $\omega(\sqrt{n\log T(n)})$ 
  bits both reliably and covertly using one of the $T(n)$ $n$-symbol slots.
Alice attempts to send one of $2^M$ (equally likely) $M$-bit 
  messages reliably to Bob using a sequence of $n$ consecutive symbol periods 
  out of $nT(n)$, where $M=\omega(\sqrt{n\log T(n)})$, and each message 
  is encoded arbitrarily into $n$ symbols.
Unlike in the previous section, here Willie is oblivious to the locations of 
  the slot boundaries, Alice's codebook construction scheme and other properties
  of her signal.
Nevertheless, by dividing his sequence of $nT(n)$ observations 
  into a set of $T(n)$ non-overlapping subsequences each containing $n$ symbols,
  and employing a simple 
  threshold detector on the maximum subsequence power, Willie can detect Alice 
  if she attempts to transmit $\omega(\sqrt{n\log T(n)})$ covert bits reliably.

\setcounter{theorem}{1}
\begin{theorem}
\label{th:converse}
Suppose Alice's transmitter is subject to the average power constraint
  $P_{\max}\in(0,\infty)$.
If Alice attempts to transmit $\omega(\sqrt{n \log T(n)})$ bits using
  a sequence of $n$ consecutive symbol periods that are arbitrarily located
  inside a sequence of $nT(n)$ symbol periods, then, as $n\rightarrow\infty$, 
  either Willie detects her with high probability, or Bob cannot decode
  with arbitrarily low probability of error.
\end{theorem}

\begin{IEEEproof}
First, consider $\log T(n)=\omega(n)$.
By the standard arguments \cite[Ch.~9]{cover02IT}, the average power constraint
  implies that Alice can reliably transmit at
  most $\mathcal{O}(n)$ bits in $n$ channel uses.
Since $n$ is asymptotically smaller than $\sqrt{n\log T(n)}$, the claim holds 
  trivially.
Therefore, we focus on $\log T(n)=\mathcal{O}(n)$ for the remainder of 
  the proof.  

Willie divides the sequence $\mathbf{Y}_{\rm w}$ of $nT(n)$ observations
  into a set of $T(n)$ non-overlapping subsequences 
  $\{\mathbf{Y}_{\rm w}(t)\}_{t=1}^{T(n)}$, with each $\mathbf{Y}_{\rm w}(t)$ containing 
  $n$ consecutive observations.
Denote by $\Upsilon_t=\sum_{Y_i\in \mathbf{Y}_{\rm w}(t)}Y_i^2$ the observed power in each
  subsequence and $\Upsilon_{\max}=\max_{t\in\{1,\ldots,T(n)\}}\Upsilon_t$.
For a threshold $\tau$, Willie accuses Alice of transmitting if $\Upsilon_{\max}>\tau$,
  setting
\begin{align}
\label{eq:tau}\tau&=\sigma_{\rm w}^2(n+c\sqrt{n\log T(n)}),
\end{align}
with constant $c>0$ determined next.

Suppose Alice does not transmit.
For an arbitrary $\mathbb{P}_{\rm FA}^\ast>0$, we show that there exists $c>0$ 
  such that the probability of false alarm
  $\mathbb{P}(\Upsilon_{\max}>\tau)\leq \mathbb{P}_{\rm FA}^*$  
   as $n\rightarrow\infty$.
Note that each $\Upsilon_t=\sigma_{\rm w}^2X_t$ where
  $\{X_t\}$, $X_t\sim\chi^2_n$, $t=1,\ldots,T(n)$ is a sequence of i.i.d.~chi-squared random variables
  each with $n$ degrees of freedom.
We have:
\begin{align*}
\mathbb{P}[\Upsilon_{\max}>\tau]&=1-\mathbb{P}\left[X_{\max}\leq \tau/\sigma_{\rm w}^2\right]=1-\left[1-\mathbb{P}\left[X_1>n+c\sqrt{n\log T(n)}\right]\right]^{T(n)},
\end{align*}
where $X_{\max}=\max_{t\in\{1,\ldots,T\}}X_t$.
The Chernoff bound for the tail of a chi-squared distribution
  \cite[Lemma 2.2]{dasgupta03jl} yields:
\begin{align}
\label{eq:chernoff_approx}\mathbb{P}[\Upsilon_{\max}>\tau]&\leq 1-\left[1-e^{\frac{n}{2}\log\left(1+\frac{c\log T(n)}{\sqrt{n}}\right)-\frac{c\sqrt{n\log T(n)}}{2}}\right]^{T(n)}\leq 1-\left[1-e^{-\frac{c^2\log T(n)}{4\left(1+c\sqrt{\frac{\log T(n)}{n}}\right)}}\right]^{T(n)}\\
\label{eq:chernoff_approx2}&=1-\left[1-\frac{1}{T(n)^{\frac{c^2}{4\left(1+c\sqrt{\frac{\log T(n)}{n}}\right)}}}\right]^{T(n)}
\end{align}
where \eqref{eq:chernoff_approx} is the application of
  $\log(1+x)\leq \frac{x(2+x)}{2(1+x)}$
  for $x\geq 0$ \cite[Eq.~(3)]{topsoe04logbounds} and re-arrangement of
  terms.
By the definition of the asymptotic notation \cite[Ch.~3.1]{clrs2e}, when 
  $\log T(n)=\mathcal{O}(n)$, there exist real constant $k>0$ and integer 
  $n_0>0$ such that $\log T(n)\leq kn$ for all $n\geq n_0$.
Setting $c>2(\sqrt{k}+\sqrt{1+k})$ ensures that \eqref{eq:chernoff_approx2}
  converges to zero as $n\rightarrow\infty$ (when $\log T(n)=o(n)$, 
  setting $c>2$ suffices).
Therefore, for any desired upper bound on the false alarm probability,
  there exists a constant $c$ such that setting the threshold as in 
  \eqref{eq:tau} guarantees that upper bound for $n$ large enough.

Now suppose Alice uses an arbitrary
  codebook $\{\mathbf{c}(W_k),k=1,\ldots,2^{nR}\}$ and transmits codeword 
  $\mathbf{c}(W_k)$ using $n$ consecutive symbol periods.
Denote the average symbol power of $\mathbf{c}(W_k)$ by
  $P_{\rm f}=\frac{\|\mathbf{c}(W_k)\|^2}{n}$.
Since Alice uses $n$ consecutive symbols, her transmission overlaps at most
  two of Willie's subsequences, which we denote $t_{\rm A}$ and $t_{\rm B}$.
Denote by $P_{\rm A}$ and $P_{\rm B}$ the power from Alice's transmission in subsequences 
  $t_{\rm A}$ and $t_{\rm B}$, respectively, with $P_{\rm A}+P_{\rm B}=nP_{\rm f}$.
Willie's probability of missing Alice's transmission is:
\begin{align}
\label{eq:factor_beta}\mathbb{P}_{\rm MD}^{(k)}&=\mathbb{P}(\Upsilon_{\max}\leq \tau)=\mathbb{P}(\Upsilon_{t_{\rm A}}\leq \tau)\mathbb{P}(\Upsilon_{t_{\rm B}}\leq \tau)\prod_{\myatop{t=1}{t\notin\{t_{\rm A},t_{\rm B}\}}}^{T(n)}\mathbb{P}(\Upsilon_t\leq \tau),
\end{align}
where the factorization in \eqref{eq:factor_beta} is because 
  Alice's codeword and the noise in other subsequences are independent.
$\prod_{t=1,t\notin\{t_{\rm A},t_{\rm B}\}}^{T(n)}\mathbb{P}(\Upsilon_t\leq \tau)\leq 1$ does not 
  depend on Alice's codeword.
However, since the codeword is an unknown deterministic signal that is added
  to AWGN on Willie's channel to Alice, 
  $\frac{\Upsilon_{t_{\rm A}}}{\sigma_{\rm w}^2}\sim\chi^2_n(P_{\rm A})$ and
  $\frac{\Upsilon_{t_{\rm B}}}{\sigma_{\rm w}^2}\sim\chi^2_n(P_{\rm B})$ are 
  non-central chi-squared random variables with $n$ degrees of freedom and
  respective non-centrality parameters $\frac{P_{\rm A}}{\sigma_{\rm w}^2}$ and 
  $\frac{P_{\rm B}}{\sigma_{\rm w}^2}$.
Without loss of generality, assume that $P_{\rm A}\geq P_{\rm B}$.
Thus, $P_{\rm A}$ satisfies $\frac{nP_{\rm f}}{2}\leq P_{\rm A}\leq nP_{\rm f}$ and the expected 
  value and variance of $\Upsilon_{t_{\rm A}}$ are bounded as 
  follows~\cite[App.~D.1]{torrieri05spreadspectrum}:
\begin{align}
\label{eq:p1_mean}\mathbb{E}\left[\Upsilon_{t_{\rm A}}\right]&\geq\sigma_{\rm w}^2n+\frac{nP_{\rm f}}{2}\\
\label{eq:p1_var}\Var\left[\Upsilon_{t_{\rm A}}\right]&\leq2n\sigma_{\rm w}^4+4n\sigma_{\rm w}^2P_{\rm f}.
\end{align}
Since $\mathbb{P}(\Upsilon_{t_{\rm B}}\leq \tau)\leq 1$, Chebyshev's inequality with 
  \eqref{eq:p1_mean} and \eqref{eq:p1_var} yields:
\begin{align}
\mathbb{P}_{\rm MD}^{(k)}&\leq\mathbb{P}\left[\left|\Upsilon_{t_{\rm A}}-\mathbb{E}[\Upsilon_{t_{\rm A}}]\right|>\mathbb{E}[\Upsilon_{t_{\rm A}}]-\sigma_{\rm w}^2[n+c\sqrt{n\log T(n)}]\right]\leq\frac{2\sigma_{\rm w}^4+4\sigma_{\rm w}^2P_{\rm f}}{\left(\frac{\sqrt{n}P_{\rm f}}{2}-c\sigma_{\rm w}^2\sqrt{\log T(n)}\right)^2}.
\end{align}
Therefore, if $P_{\rm f}=\omega\left(\sqrt{\frac{\log T(n)}{n}}\right)$,
  as $n\rightarrow \infty$, $\mathbb{P}_{\rm FA}+\mathbb{P}_{\rm MD}$ can be 
  made arbitrarily small.
The proof of the non-zero lower bound on the decoding error probability
  $\mathbb{P}_{\rm e}^{({\rm b})}$ if Alice tries to transmit $\omega(\sqrt{n\log T(n)})$ 
  bits in a single slot using
  average symbol power $P_{\rm f}=\mathcal{O}\left(\sqrt{\frac{\log T(n)}{n}}\right)$
  follows from the arguments in the proof of Theorem 2 in 
  \cite{bash12sqrtlawisit,bash13squarerootjsac}.
\end{IEEEproof}

\section{Discussion}
\label{sec:discussion}

Here we relate our work to other studies of covert communication.
An overview of the area can be found in \cite{bash15covertcommmag}.

\subsection{Relationship with Steganography}
Steganography is an ancient discipline \cite{herodotus} of hiding 
  messages in innocuous objects.
Modern steganographic systems \cite{fridrich09stego} hide information 
  by altering the properties
  of fixed-size, finite-alphabet covertext objects (e.g., images),
  and are subject to a similar SRL as covert communication:
  $\mathcal{O}(\sqrt{n})$ symbols in covertext of size $n$ may safely be 
  modified to hide an $\mathcal{O}(\sqrt{n}\log n)$-bit message 
  in the resulting stegotext \cite{ker10sqrtlawnokey}.
The similarity between the SRLs in these disciplines comes from the 
  mathematics of statistical hypothesis testing, as discussed in
  \cite{bash13squarerootjsac}.
The extra $\log n$ factor is because of the lack of noise in
  the steganography context.
However, arguably the earliest work on SRL \cite{korzhik05srl} shows its 
  achievability without the $\log n$ factor in the presence of an ``active''
  adversary that corrupts stegotext using AWGN.
This was re-discovered independently and published with the converse in
  \cite{bash12sqrtlawisit,bash13squarerootjsac}.

Batch steganography uses multiple covertext objects to hide a message
  and is subject to the steganographic SRL described above
  \cite{ker07sqrtlaw,ker07pool}.
The batch steganography interpretation of covert communication using the 
  timing-based degree-of-freedom that is
  described here is equivalent to using only one of 
  $T(n)$ covertext objects of size $n$ to embed a message.
Willie, who knows that one covertext object is used but not which one, has
  to  examine all of them.
We are not aware of any work on this particular problem, but it is likely that
  one could extend our result to it.
We also note that more recent work on steganography shows that an empirical 
  model of the covertext suffices to break the steganographic SRL,
  allowing the embedding of $\mathcal{O}(n)$ bits in an $n$-symbol covertext
  \cite{craver10nosqrtlaw}.
However, this technique relies on embedding messages in covertext by
  \emph{replacing} part of it---something that cannot be done in standard
  communication systems unless Alice controls Willie's noise source.

\subsection{Related Work in Physical Layer Covert Communication}
The emergence of radio-frequency (RF) communication systems necessitated 
  the development of means to protect them from jamming, detection, and 
  eavesdropping.
Spread-spectrum techniques \cite{simon94ssh} address these issues by 
  transmitting a signal that requires bandwidth $W_{\rm M}$ on a much wider 
  bandwidth $W_{\rm s}\gg W_{\rm M}$,
  thus, effectively suppressing the power spectral density of the signal below 
  the noise floor.
This provides both covertness as well as the resistance to jamming, fading,
  and other interference.

However, while the spread-spectrum architectures are well-developed, 
  the fundamental SRL for covert communication has been derived only recently
  \cite{bash12sqrtlawisit,bash13squarerootjsac}.
This resulted in the revival of the field, with follow-on work focusing
  on reducing the size of the pre-shared secret
  \cite{che13sqrtlawbscisit, bloch15covert}, fully characterizing 
  the optimal constant hidden by the big-$\mathcal{O}$ notation of the SRL
  \cite{bloch15covert, wang15covert}, and extending the SRL to 
  quantum channels with Willie limited only by the laws of quantum mechanics
  \cite{bash13quantumlpdisit,bash15covertbosoniccomm,azadeh15quantumcovertarxiv}.
Finally, while here we improve on the SRL by exploiting Willie's ignorance of
  transmission timing, other studies explore even stronger 
  assumptions on his limitations.
In particular, authors in 
  \cite{lee15posratecovertjstsp,che14channeluncertainty,sobers15jammer-asilomar}
  examine the impact of the errors in Willie's estimate of noise
  variance $\sigma_{\rm w}^2$ at his receiver,
  which allows $\mathcal{O}(n)$ covert
  bits to be transmitted in $n$ uses of the channel even when Willie has
  upper and lower bounds on $\sigma_{\rm w}^2$.
Thus, positive-rate rather than SRL-governed covert communication is possible
  when Willie's knowledge of the channel is incomplete.
However, successive work has demonstrated that the converse of Section 
  \ref{sec:converse} still holds even if Willie does not know 
  $\sigma_{\rm w}^2$, but under the restrictive 
  assumption that Willie knows the slot boundaries\cite{goeckel16timing}.  
The main result of \cite{goeckel16timing} can then be combined with 
  the approach of Theorem \ref{th:converse} to remove the requirement of 
  Willie knowing the slot boundaries, as described in Corollary 2 of 
  \cite{goeckel16timing}.

\section{Conclusion and Future Work}
\label{sec:conclusion}
We have shown that secretly pre-arranging a choice of a single $n$-symbol period
  slot out of $T(n)$ allows Alice to reliably transmit 
  $\mathcal{O}(\min\{n,\sqrt{n\log T(n)}\})$ bits on an AWGN channel to Bob 
  while rendering Willie's detector arbitrarily close to ineffective.
Surprisingly, the multiplicative increase in transmitted information 
  over the result in 
  \cite{bash12sqrtlawisit,bash13squarerootjsac} is obtained without the need
  for Bob to know which slot holds the transmission if $T(n)<2^{c_{\rm T}n}$, 
  where $c_{\rm T}$ is a constant, and, when $T(n)\geq 2^{c_{\rm T}n}$ only an 
  additive expense of an extra $\log T(n)$ pre-shared secret bits is needed.
In the future we plan on combining this work with our recent results on 
  jammer-assisted covert communication \cite{soltani14netlpdallerton} to
  enable covert networks.

\appendices

\setcounter{lemma}{0}
\section{Proof of Lemma \ref{lemma:LRT_bound}}
\label{app:convergence}
\begin{figure}[h]
\vspace{-0.05in}
\begin{center}
\begin{picture}(85,20)
\put(66,12){\vector(1,0){19}}
\put(25,12){\vector(-1,0){25}}
\put(60,12){\line(-1,0){29}}
\put(61,9.90){\makebox(4,4){$\ldots$}}
\put(26,9.90){\makebox(4,4){$\ldots$}}
\put(26.2,7){\makebox(4,4){\scriptsize $0$}}
\put(0,2.5){\makebox(10,4){\scriptsize $\leftarrow$~Region 1}}
\put(13,4){\line(0,1){3}}
\put(72,4){\line(0,1){3}}
\put(0,7){\line(1,0){85}}
\put(75,2.5){\makebox(10,4){\scriptsize Region 3~$\rightarrow$}}
\put(37,2.5){\makebox(10,4){\scriptsize Region 2}}
\put(13,10.5){\line(0,1){3}}
\put(72,10.5){\line(0,1){3}}
\put(11,7){\makebox(4,4){\scriptsize $G$}}
\put(70,7){\makebox(4,4){\scriptsize $H$}}
\put(8,11){\line(0,1){2}}
\put(5,7){\makebox(5,4){\tiny $G\hspace{-.36mm}-\hspace{-.36mm}\delta$}}
\put(18,11){\line(0,1){2}}
\put(15,7){\makebox(5,4){\tiny $G\hspace{-.36mm}+\hspace{-.36mm}\delta$}}
\put(23,11){\line(0,1){2}}
\put(67,11){\line(0,1){2}}
\put(64,7){\makebox(5,4){\tiny $H\hspace{-.36mm}-\hspace{-.36mm}\delta$}}
\put(77,11){\line(0,1){2}}
\put(74,7){\makebox(5,4){\tiny $H\hspace{-.36mm}+\hspace{-.36mm}\delta$}}
\put(54,15.5){\makebox(8,4){\tiny $\tau(n)\hspace{-.36mm}+\hspace{-.36mm}\delta$}}
\put(46,15.5){\makebox(8,4){\tiny $\tau(n)$}}
\put(38,15.5){\makebox(8,4){\tiny $\tau(n)\hspace{-.36mm}-\hspace{-.36mm}\delta$}}
\put(45,15){\line(1,0){10}}
\put(45,14.5){\line(0,1){1}}
\put(50,14){\line(0,1){2}}
\put(55,14.5){\line(0,1){1}}
\put(32,11){\line(0,1){2}}
\put(37,11){\line(0,1){2}}
\put(42,11){\line(0,1){2}}
\put(40,7){\makebox(4,4){\tiny $x_k$}}
\put(47,11){\line(0,1){2}}
\put(52,11){\line(0,1){2}}
\put(57,11){\line(0,1){2}}
\put(54.7,7){\makebox(4,4){\tiny $x_k+3\delta$}}
\end{picture}
\end{center}
\vspace{-0.20in}
\caption{The real number line partitioned into three regions  
  for the analysis of $\mathbb{P}(E_{\rm C}(\tau(n),\delta))$.
  $G$, $H$ and $\delta$ are the constants that we select. $\tau(n)$ 
  satisfying $G\leq \tau(n)\leq H$ is illustrated.}
\label{fig:regions}
\vspace{-0.05in}
\end{figure}
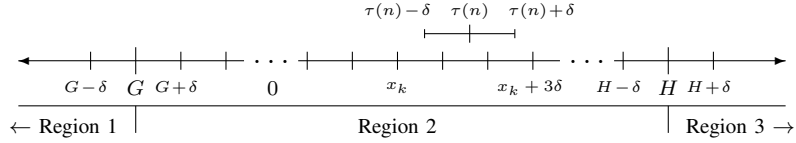

Consider any $\epsilon>0$.
Suppose Willie chooses threshold $\tau(n)$ arbitrarily.
The false alarm probability is lower-bounded using the fact that $S^{(n)}$ is 
  independent of which hypothesis is true:
\begin{align}
\mathbb{P}[L^{(n)}_0>\tau(n)|H_0\text{~is true}]&\geq\mathbb{P}\left[S^{(n)}\geq \tau(n)+\delta\big||V^{(n)}_0|<\delta\right].\nonumber
\end{align}
Similarly the probability of missed detection is lower-bounded as follows:
\begin{align}
\mathbb{P}[L^{(n)}_1\leq\tau(n)|H_1\text{~is true}]&\geq\mathbb{P}\left[S^{(n)}\leq \tau(n)-\delta\big||V^{(n)}_1|<\delta\right].\nonumber
\end{align}
Denoting by $E_{\rm C}(\tau(n),\delta)$ the event that either
  $S^{(n)}\geq \tau(n)+\delta$ or $S^{(n)}\leq \tau(n)-\delta$,
\begin{align}
\mathbb{P}(E_{\rm C}(\tau(n),\delta))&=1-F_{S^{(n)}}(\tau(n)+\delta)+F_{S^{(n)}}(\tau(n)-\delta),\nonumber
\end{align}
where $F_{S^{(n)}}(\cdot)$ is the distribution function for $S^{(n)}$.
Denote the standard Gaussian distribution function by 
  $\Phi(z)=\int_{-\infty}^z\phi(t)\mathrm{d}t$ where 
  $\phi(t)=\frac{e^{-t^2/2}}{\sqrt{2\pi}}$ is the standard Gaussian density 
  function.
The convergence of $F_{S^{(n)}}(z)$ to $\Phi(z)$ is pointwise in
  $z$, and, since $\tau(n)$ is the $n^{\text{th}}$ value in an 
  arbitrary sequence, we cannot use this fact directly.
However, let's choose finite constants $G<0$ and $H>0$, and partition the real 
  number line into three regions as shown in Figure \ref{fig:regions}.
Clearly, for any $n$, $\tau(n)$ is in one of these regions.
Next we demonstrate that \eqref{eq:phi} holds for an arbitrary $\tau(n)$
  by appropriately selecting $G$, $H$, and $\delta$.

Consider $\tau(n)<G$, or region 1 in Figure \ref{fig:regions}:
$\mathbb{P}(E_{\rm C}(\tau(n),\delta))\geq 1-F_{S^{(n)}}(\tau(n)+\delta)\geq 1-F_{S^{(n)}}(G+\delta)$.
Because the convergence of $F_{S^{(n)}}(z)$ to $\Phi(z)$ is pointwise, given 
  $\delta$, $\epsilon$, and $G=\Phi^{-1}(\epsilon/6)-\delta$, there 
  exists $n_2$ such that, for all $n\geq n_2$,
 $\mathbb{P}(E_{\rm C}(\tau(n),\delta))\geq1-\Phi(G+\delta)-\frac{\epsilon}{6}=1-\frac{\epsilon}{3}$ 
when $\tau(n)<G$.
Similarly for $\tau(n)>H$, or region 3 in Figure \ref{fig:regions}:
$\mathbb{P}(E_{\rm C}(\tau(n),\delta))\geq F_{S^{(n)}}(\tau(n)-\delta)\geq F_{S^{(n)}}(H+\delta)$.
Again, because the convergence of $F_{S^{(n)}}(z)$ to $\Phi(z)$ is pointwise, 
  given $\delta$, $\epsilon$, and $H=\Phi^{-1}(1-\epsilon/6)+\delta$, there 
  exists $n_3$ such that, for $\tau(n)>H$ and all $n\geq n_3$,
\begin{align}
\label{eq:H}\mathbb{P}(E_{\rm C}(\tau(n),\delta))&\geq\Phi(H+\delta)-\frac{\epsilon}{3}=1-\frac{\epsilon}{3}.
\end{align}

Finally, consider $\tau(n)$ satisfying $G\leq \tau(n)\leq H$, or region 2 
  in Figure \ref{fig:regions}.
Let's assume that $H$ and $G$ are selected so that $H-G$ is an integer multiple
  of $\delta$ (e.g., using larger $H$ than necessary, 
  which results in the lower bound in \eqref{eq:H} being smaller).
Consider a sequence 
  $\{x_k\}_{k=0}^{(H-G)/\delta+2}$ where $x_0=G-\delta$, 
  $x_1=G$, $x_2=G+\delta$,
  $x_3=G+2\delta,\ldots,x_{(H-G)/\delta}=H-\delta$, $x_{(H-G)/\delta+1}=H$,
  $x_{(H-G)/\delta+2}=H+\delta$.
Sequence $\{x_k\}_{k=0}^{(H-G)/\delta+2}$
  partitions region 2 into $\frac{H-G}{\delta}+2$ subregions, and,
  for any $\tau(n)$ satisfying $G\leq \tau(n)\leq H$, there exists 
  $k\in\left\{0,\ldots,\frac{H-G}{\delta}+2\right\}$ such that 
  $x_k\leq \tau(n)-\delta<\tau(n)+\delta\leq x_k+3\delta$, as
  illustrated in Figure \ref{fig:regions}.
Since $F_{S^{(n)}}(z)$ is monotonic,
\begin{align}
\mathbb{P}(E_{\rm C}(\tau(n),\delta))&\geq1-F_{S^{(n)}}(x_k+3\delta)+F_{Z(n)}(x_k).
\end{align}
Since the convergence of $F_{S^{(n)}}(z)$ to $\Phi(z)$ is pointwise, for a given
  $x_k$, $\delta$, and $\epsilon$, there exists $m_k$ such that for all 
  $n\geq m_k$,
\begin{align}
\mathbb{P}(E_{\rm C}(\tau(n),\delta))&\geq1-\left(\Phi(x_k+3\delta)+\frac{\epsilon}{12}\right)+\left(\Phi(x_k)-\frac{\epsilon}{12}\right)=1-\int_{x_k}^{x_k+3\delta}\phi(t)\mathrm{d}t-\frac{\epsilon}{6}\\
\label{eq:normpdfmax}&\geq1-\frac{3\delta}{\sqrt{2\pi}}-\frac{\epsilon}{6}
\end{align}
where \eqref{eq:normpdfmax} follows from $\phi(t)\leq\frac{1}{\sqrt{2\pi}}$.
Setting $\delta=\frac{\epsilon\sqrt{2\pi}}{18}$ and 
  $n_4=\max_{\left\{0,\ldots,\frac{H-G}{\delta}\right\}}(m_k)$ yields 
  the desired lower bound for all $n\geq n_4$ when $\tau(n)$ satisfies 
  $G\leq \tau(n)\leq H$.
Thus, for any $\tau(n)$, when $n\geq n_0$ where $n_0=\max(n_2,n_3,n_4)$,
\begin{align}
\label{eq:phi}\mathbb{P}\left(E_{\rm C}\left(\tau(n),\epsilon\sqrt{2\pi}/18\right)\right)&\geq 1-\frac{\epsilon}{3}.
\end{align}

Since $V^{(n)}_0\xrightarrow{\mathcal{P}}0$ and 
  $V^{(n)}_1\xrightarrow{\mathcal{P}}0$, there exists $n_1>0$ such that
  for all $n\geq n_1$,
  $\mathbb{P}\left(|V^{(n)}_0|>\frac{\epsilon\sqrt{2\pi}}{18}\right)<\frac{\epsilon}{3}$
  and
  $\mathbb{P}\left(|V^{(n)}_0|>\frac{\epsilon\sqrt{2\pi}}{18}\right)<\frac{\epsilon}{3}$.
The intersection of these events and the event 
  $E_{\rm C}(S(n),\epsilon\sqrt{2\pi}/9)$ yields a detection error event.
By combining their probabilities using DeMorgan's Law and the union bound,
  we lower-bound $\mathbb{P}_{\rm FA}+\mathbb{P}_{\rm MD}\geq1-\epsilon$
  for all $n\geq \max\{n_0,n_1\}$.

\textit{Remark:} Lemma \ref{lemma:LRT_bound} holds when 
  $S^{(n)}$ converges to any distribution provided it has a continuous 
  density, however, here we do not need such generality.

\section{Proof of equality $\frac{1}{2^n}\sum_{\mathbf{x}\in\{-1,1\}^n}\exp\left[a\sum_{i=1}^n x_i\right]=\cosh^n(a)$}
\label{app:derive_cosh}

We argue by induction on $n$.
The base case $n=1$ is trivial.
Assume the claim holds for $n$.
Now,
\begin{align*}
\frac{1}{2^{n+1}}\sum_{\mathbf{x}\in\{-1,1\}^{n+1}}\exp\left[a\sum_{i=1}^{n+1} x_i\right]&=\frac{1}{2\cdot2^{n}}\sum_{\myatop{\mathbf{x}\in\{-1,1\}^{n}}{x_{n+1}\in\{-1,1\}}}\exp[ax_{n+1}]\exp\left[a\sum_{i=1}^{n} x_i\right]=\cosh(a)\cosh^n(a).
\end{align*}

\section{Analysis of $\mathbb{P}_{\rm e}^{({\rm b})}$ under peak power constraint and known $t_{\rm A}$}
\label{app:corr}
Here we analyze Bob's decoding error probability $\mathbb{P}_{\rm e}^{({\rm b})}$ when 
  the transmission time $t_{\rm A}$ is known and Alice uses binary modulation
  $\{-a,a\}$ that satisfies the peak power constraint and ensures that
  $\mathbb{P}_{\rm FA}+\mathbb{P}_{\rm MD}\geq1-\epsilon$.
This appendix provides an alternative to the analysis of $\mathbb{P}_{\rm e}^{({\rm b})}$ in
  the proof of Theorem~1.2 in \cite{bash12sqrtlawisit,bash13squarerootjsac} 
  (the analysis of $\mathbb{P}_{\rm e}^{({\rm b})}$ in the proof of Theorem~1.2
  in \cite{bash12sqrtlawisit,bash13squarerootjsac} contains minor technical 
  errors which do \emph{not} change the main results).
The construct presented here is adapted for the proof of Theorem 
  \ref{th:achievability_peak_power} in Section \ref{sec:achievability}.

Suppose Alice transmits $\mathbf{c}(W_k)$.
Recall that $\mathbf{c}(W_k)$ is drawn from a codebook 
  $\{\mathbf{c}(W_m),m=1,2,\ldots,2^M\}$ containing codewords that are 
  independently generated according to 
  $p_{\mathbf{X}}(\mathbf{x})=\prod_{i=1}^n p_X(x_i)$, where
  $p_X(-a)=p_X(a)=\frac{1}{2}$.
Bob uses an ML decoder which suffers an error event $E_{k\rightarrow i}$ when
  the received vector $\mathbf{Y}_b$ is closer in Euclidean
  distance to codeword $\mathbf{c}(W_i)$, $i\neq k$.
The decoding error probability, averaged over all the codebooks, is then:
\begin{align}
\mathbb{P}_{e}^{({\rm b})}&=\mathbb{E}_{\mathbf{c}(W_k)}\left[\mathbb{P}\left(\cup_{\myatop{i=0}{i\neq k}}^{2^M}E_{k\rightarrow i}\right)\right]\leq\sum_{\myatop{i=1}{i\neq k}}^{2^M}\mathbb{E}_{\mathbf{c}(W_k)}\mathbb{P}(E_{k\rightarrow i})\label{eq:av_bincodebook_err},
\end{align}
where $\mathbb{E}_X[\cdot]$ denotes the expectation over random variable $X$,
  and the inequality in \eqref{eq:av_bincodebook_err} is the union bound.
Let $\|\mathbf{d}\|_2=\|\mathbf{c}(W_k)-\mathbf{c}(W_i)\|_2$ denote 
  the Euclidean ($\mathcal{L}_2$) distance between two codewords.
Then, by \cite[Eq.~(3.44)]{madhow08digicom}:
\begin{align}
\mathbb{E}_{\mathbf{c}(W_k)}\mathbb{P}(E_{k\rightarrow i})&=\mathbb{E}_{\mathbf{d}}\left[Q\left(\frac{\|\mathbf{d}\|_2}{2\sigma_{\rm b}}\right)\right]\label{eq:Qbound}\leq\mathbb{E}_{\mathbf{d}}\left[\frac{1}{2}\exp\left(-\frac{\|\mathbf{d}\|_2^2}{8\sigma_{\rm b}^2}\right)\right],
\end{align}
where $Q(x)=\frac{1}{\sqrt{2\pi}}\int_{x}^{\infty}e^{-t^2/2}\mathrm{d}t$ and 
  the bound in \eqref{eq:Qbound} is because  
  $Q(x)\leq\frac{1}{2}e^{-x^2/2}$ \cite[Eq.~(5)]{chiani03qfunction}.
The Euclidean distance $\|\mathbf{d}\|_2$ depends on the number of locations $j$
  where $\mathbf{c}(W_k)$ and $\mathbf{c}(W_i)$ differ, which is 
  binomially-distributed by construction, where each location is different
  with probability $\frac{1}{2}$.
When the codewords are different in $j$ locations, we have
  $\|\mathbf{d}\|_2^2=4ja^2$, and thus:
\begin{align}
\mathbb{E}_{\mathbf{d}}\left[\frac{1}{2}\exp\left(-\frac{\|\mathbf{d}\|_2^2}{8\sigma_{\rm b}^2}\right)\right]&=\frac{1}{2}\sum_{j=0}^n \exp\left(-\frac{ja^2}{2\sigma_{\rm b}^2}\right)\binom{n}{j}\frac{1}{2^n}=\frac{1}{2^{n+1}}\left[1+\exp\left(-\frac{a^2}{2\sigma_{\rm b}^2}\right)\right]^n\label{eq:exp_bincodeword_err},
\end{align}
where \eqref{eq:exp_bincodeword_err} is an application of binomial theorem.
Substitution of \eqref{eq:exp_bincodeword_err} into 
  \eqref{eq:av_bincodebook_err} yields
\begin{align}
\label{eq:p_e_b_ub}\mathbb{P}_{e}^{({\rm b})}&\leq 2^{M-n\left(1-\log_2\left[1+\exp\left(-\frac{a^2}{2\sigma_{\rm b}^2}\right)\right]\right)}.
\end{align}
Thus, if Alice attempts to transmit 
  $M=n\gamma\left(1-\log_2\left[1+\exp\left(-\frac{a^2}{2\sigma_{\rm b}^2}\right)\right]\right)$
  bits, where $\gamma\in(0,1)$ is a constant, Bob's decoding error probability
  decays to zero as $n\rightarrow\infty$.
If $a^2=\mathcal{O}(1)$, then clearly $M=\mathcal{O}(n)$ bits. 
If $a^2=o(1)$, then, application of the bounds $\log_2(1+x)\leq \frac{x}{\ln 2}$
  and $1-e^{-x}\geq x-\frac{x^2}{2}$ yields $M=\mathcal{O}(na^2)$.
In particular, if $\sigma_{\rm w}^2$ is known to Alice and
  $a^2=\frac{2\sqrt{2}\epsilon \sigma_{\rm w}^2}{\sqrt{n}}$ as prescribed by 
  the analysis of Willie's detector in \cite[Theorem 1.2]{bash13squarerootjsac},
  $M=\mathcal{O}(\epsilon\sigma_{\rm w}^2\sqrt{n}/\sigma_{\rm b}^2)$.

\bibliographystyle{IEEEtran}

\begin{thebibliography}{10}
\providecommand{\url}[1]{#1}
\csname url@samestyle\endcsname
\providecommand{\newblock}{\relax}
\providecommand{\bibinfo}[2]{#2}
\providecommand{\BIBentrySTDinterwordspacing}{\spaceskip=0pt\relax}
\providecommand{\BIBentryALTinterwordstretchfactor}{4}
\providecommand{\BIBentryALTinterwordspacing}{\spaceskip=\fontdimen2\font plus
\BIBentryALTinterwordstretchfactor\fontdimen3\font minus
  \fontdimen4\font\relax}
\providecommand{\BIBforeignlanguage}[2]{{%
\expandafter\ifx\csname l@#1\endcsname\relax
\typeout{** WARNING: IEEEtran.bst: No hyphenation pattern has been}%
\typeout{** loaded for the language `#1'. Using the pattern for}%
\typeout{** the default language instead.}%
\else
\language=\csname l@#1\endcsname
\fi
#2}}
\providecommand{\BIBdecl}{\relax}
\BIBdecl

\bibitem{bash14timingisit}
B.~A. Bash, D.~Goeckel, and D.~Towsley, ``{LPD Communication when the Warden
  Does Not Know When},'' in \emph{Proc. {IEEE} Int.~Symp.~Inform.~Theory
  ({ISIT})}, Honolulu, HI, Jul. 2014.

\bibitem{bbc14snowden}
{BBC}, ``{Edward Snowden: Leaks that exposed US spy programme},''
  \url{http://www.bbc.com/news/world-us-canada-23123964}, Jan. 2014.

\bibitem{bash12sqrtlawisit}
B.~A. Bash, D.~Goeckel, and D.~Towsley, ``Square root law for communication
  with low probability of detection on {AWGN} channels,'' in \emph{Proc. {IEEE}
  Int. Symp. Inform. Theory ({ISIT})}, Cambridge, MA, Jul. 2012.

\bibitem{bash13squarerootjsac}
------, ``Limits of reliable communication with low probability of detection on
  {AWGN} channels,'' \emph{{IEEE} J. Select. Areas Commun.}, vol.~31, no.~9,
  pp. 1921--1930, 2013, arXiv:1202.6423.

\bibitem{che13sqrtlawbscisit}
P.~H. Che, M.~Bakshi, and S.~Jaggi, ``Reliable deniable communication: Hiding
  messages in noise,'' in \emph{Proc. {IEEE} Int. Symp. Inform. Theory
  ({ISIT})}, Istanbul, Turkey, Jul. 2013, arXiv:1304.6693.

\bibitem{bloch15covert}
M.~R. Bloch, ``Covert communication over noisy channels: A resolvability
  perspective,'' \emph{IEEE Trans. Inf. Theory}, vol.~62, no.~5, pp.
  2334--2354, May 2016.

\bibitem{wang15covert}
L.~Wang, G.~W. Wornell, and L.~Zheng, ``Fundamental limits of communication
  with low probability of detection,'' \emph{IEEE Trans. Inf. Theory}, vol.~62,
  no.~6, pp. 3493--3503, June 2016.

\bibitem{kadhe14sqrtlawmultipathisit}
S.~Kadhe, S.~Jaggi, M.~Bakshi, and A.~Sprintson, ``Reliable, deniable, and
  hidable communication over multipath networks,'' in \emph{Proc. {IEEE} Int.
  Symp. Inform. Theory ({ISIT})}, Honolulu, HI, Jul. 2014, arXiv:1401.4451.

\bibitem{soltani14netlpdallerton}
R.~Soltani, B.~A. Bash, D.~Goeckel, S.~Guha, and D.~Towsley, ``Covert
  single-hop communication in a wireless network with distributed artificial
  noise generation,'' in \emph{Proc. Conf. Commun. Control Comp. (Allerton)},
  Monticello, {IL}, 2014.

\bibitem{bash13quantumlpdisit}
B.~A. Bash, S.~Guha, D.~Goeckel, and D.~Towsley, ``{Quantum Noise Limited
  Communication with Low Probability of Detection},'' in \emph{Proc. {IEEE}
  Int. Symp. Inform. Theory ({ISIT})}, Istanbul, Turkey, Jul. 2013.

\bibitem{bash15covertbosoniccomm}
B.~A. Bash, A.~H. Gheorghe, M.~Patel, J.~L. Habif, D.~Goeckel, D.~Towsley, and
  S.~Guha, ``Quantum-secure covert communication on bosonic channels,''
  \emph{Nat Commun}, vol.~6, Oct 2015.

\bibitem{azadeh15quantumcovertarxiv}
A.~Sheikholeslami, B.~A. Bash, D.~Towsley, D.~Goeckel, and S.~Guha, ``Covert
  communication over classical-quantum channels,'' arXiv:1601.06826 [quant-ph],
  2016.

\bibitem{lee15posratecovertjstsp}
S.~Lee, R.~Baxley, M.~Weitnauer, and B.~Walkenhorst, ``Achieving undetectable
  communication,'' \emph{{IEEE} J. Select. Topics Signal Proc.}, vol.~9, no.~7,
  pp. 1195--1205, Oct. 2015.

\bibitem{hou14isit}
J.~Hou and G.~Kramer, ``Effective secrecy: Reliability, confusion and
  stealth,'' in \emph{Proc.~of {IEEE} Int.~Symp.~Inform.~Theory ({ISIT})},
  Honolulu, HI, Jul. 2014, arXiv:1311.1411.

\bibitem{han14reliabilitysecrecy}
T.~S. Han, H.~Endo, and M.~Sasaki, ``Reliability and secrecy functions of the
  wiretap channel under cost constraint,'' \emph{IEEE Trans. Inf. Theory},
  vol.~60, no.~11, pp. 6819--6843, Nov 2014.

\bibitem{che14channeluncertainty}
P.~H. Che, M.~Bakshi, C.~Chan, and S.~Jaggi, ``Reliable deniable communication
  with channel uncertainty,'' in \emph{{Proc. Inform. Theory Workshop (ITW)}},
  Hobart, Tasmania, Australia, Nov 2014, pp. 30--34.

\bibitem{sobers15jammer-asilomar}
T.~V. Sobers, B.~A. Bash, D.~Goeckel, S.~Guha, and D.~Towsley, ``Covert
  communication with the help of an uninformed jammer achieves positive rate,''
  in \emph{Asilomar Conf. Signals Syst. Comput.}, Nov 2015, pp. 625--629.

\bibitem{clrs2e}
T.~H. Cormen, C.~E. Leiserson, R.~L. Rivest, and C.~Stein, \emph{Introduction
  to Algorithms}, 2nd~ed.\hskip 1em plus 0.5em minus 0.4em\relax Cambridge,
  Massachusetts: {MIT} Press, 2001.

\bibitem{arumugam16async}
K.~S.~K. Arumugam and M.~R. Bloch, ``Keyless asynchronous covert
  communication,'' in \emph{Proc. Inform. Theory Workshop ({ITW})}, Sep. 2016.

\bibitem{shannon49sec}
C.~E. Shannon, ``Communication theory of secrecy systems,'' \emph{Bell System
  Technical Journal}, vol.~28, pp. 656--715, 1949.

\bibitem{menezes96HAC}
A.~J. Menezes, S.~A. Vanstone, and P.~C.~V. Oorschot, \emph{Handbook of Applied
  Cryptography}.\hskip 1em plus 0.5em minus 0.4em\relax Boca Raton, FL, USA:
  CRC Press, Inc., 1996.

\bibitem{lehmann05stathyp}
E.~Lehmann and J.~Romano, \emph{Testing Statistical Hypotheses}, 3rd~ed.\hskip
  1em plus 0.5em minus 0.4em\relax New York: Springer, 2005.

\bibitem{cover02IT}
T.~M. Cover and J.~A. Thomas, \emph{Elements of Information Theory},
  2nd~ed.\hskip 1em plus 0.5em minus 0.4em\relax John Wiley \& Sons, Hoboken,
  NJ, 2002.

\bibitem{billingsley95measure}
P.~Billingsley, \emph{{Probability and Measure}}, 3rd~ed.\hskip 1em plus 0.5em
  minus 0.4em\relax \hspace{-2pt}{New York}: Wiley, 1995.

\bibitem{boyd04convexopt}
S.~Boyd and L.~Vandenberghe, \emph{Convex Optimization}.\hskip 1em plus 0.5em
  minus 0.4em\relax New York, NY, USA: Cambridge University Press, 2004.

\bibitem{madhow08digicom}
U.~Madhow, \emph{Fundamentals of Digital Communication}.\hskip 1em plus 0.5em
  minus 0.4em\relax Cambridge, UK: Cambridge University Press, 2008.

\bibitem{chiani03qfunction}
M.~Chiani, D.~Dardari, and M.~K. Simon, ``New exponential bounds and
  approximations for the computation of error probability in fading channels,''
  \emph{{IEEE} Trans. Wireless Commun.}, vol.~2, no.~4, pp. 840--845, Jul.
  2003.

\bibitem{dasgupta03jl}
S.~Dasgupta and A.~Gupta, ``{An Elementary Proof of a Theorem of Johnson and
  Lindenstrauss},'' \emph{Random Struct. Algorithms}, vol.~22, no.~1, pp.
  60--65, Jan. 2003.

\bibitem{topsoe04logbounds}
F.~Tops{\o}e, ``Some bounds for the logarithmic function,'' \emph{{RGMIA}
  Research Rep. Collection}, vol.~7, no.~2, 2004.

\bibitem{torrieri05spreadspectrum}
D.~Torrieri, \emph{Principles of Spread-spectrum Communication Systems}.\hskip
  1em plus 0.5em minus 0.4em\relax Boston, MA, USA: Springer, 2005.

\bibitem{bash15covertcommmag}
B.~A. Bash, D.~Goeckel, S.~Guha, and D.~Towsley, ``Hiding information in noise:
  Fundamental limits of covert wireless communication,'' \emph{{IEEE} Commun.
  Mag.}, vol.~53, no.~12, 2015.

\bibitem{herodotus}
Herodotus, c.~440 BCE, 5.35 and 7.239.

\bibitem{fridrich09stego}
J.~Fridrich, \emph{Steganography in Digital Media: Principles, Algorithms, and
  Applications}, 1st~ed.\hskip 1em plus 0.5em minus 0.4em\relax New York:
  Cambridge University Press, 2009.

\bibitem{ker10sqrtlawnokey}
A.~D. Ker, ``The square root law does not require a linear key,'' in
  \emph{Proc. ACM Workshop Multimedia Security}, Roma, Italy, 2010, pp.
  213--224.

\bibitem{korzhik05srl}
V.~Korzhik, G.~Morales-Luna, and M.~H. Lee, ``On the existence of perfect
  stegosystems,'' in \emph{Proc. 4th Int. Workshop Digital Watermarking
  (IWDW)}, Siena, Italy, Sep. 2005, pp. 30--38.

\bibitem{ker07sqrtlaw}
A.~D. Ker, ``A capacity result for batch steganography,'' \emph{{IEEE} Signal
  Process. Lett.}, vol.~14, no.~8, pp. 525--528, Aug. 2007.

\bibitem{ker07pool}
------, ``Batch steganography and pooled steganalysis,'' in \emph{Proc. Int.
  Inform. Hiding Workshop}, Alexandria, VA, 2006, pp. 265--281.

\bibitem{craver10nosqrtlaw}
S.~Craver and J.~Yu, ``Subset selection circumvents the square root law,'' in
  \emph{Proc. SPIE Media Forensics Security}, San Jose, CA, 2010, pp.
  754\,103--1--754\,103--6.

\bibitem{simon94ssh}
M.~K. Simon, J.~K. Omura, R.~A. Scholtz, and B.~K. Levitt, \emph{Spread
  Spectrum Communications Handbook}.\hskip 1em plus 0.5em minus 0.4em\relax
  McGraw-Hill, 1994.

\bibitem{goeckel16timing}
D.~Goeckel, B.~Bash, S.~Guha, and D.~Towsley, ``Covert communications when the
  warden does not know the background noise power,'' \emph{IEEE Commun. Lett.},
  vol.~20, no.~2, pp. 236--239, Feb 2016.

\end{thebibliography}

\end{document}